\documentclass[pdflatex,sn-mathphys-num]{sn-jnl}

\usepackage{setspace}

\usepackage{graphicx}
\usepackage{amsmath,amssymb,amsfonts,bm}
\usepackage[title]{appendix}
\usepackage{xcolor}
\usepackage{textcomp}
\usepackage{booktabs,array}
\usepackage{tabularx}
\usepackage{siunitx}
\usepackage{microtype}
\usepackage{placeins}

\newcolumntype{L}[1]{>{\raggedright\arraybackslash}p{#1}}
\newcolumntype{Y}{>{\raggedright\arraybackslash}X}
\newcolumntype{C}{>{\centering\arraybackslash}X}
\newcolumntype{Z}[1]{>{\centering\arraybackslash}p{#1}}

\newcommand{\sech}{\operatorname{sech}}

\newcommand{\EqualGaussianMean}{884.621}
\newcommand{\EqualGaussianRMS}{9.864}
\newcommand{\WeightedMean}{878.689}
\newcommand{\InternalSE}{0.255}
\newcommand{\ChiSq}{72.875}
\newcommand{\ReducedChiSq}{3.644}
\newcommand{\PDGScale}{1.909}
\newcommand{\PDGSE}{0.486}
\newcommand{\SteinerGScale}{2.397}

\newcommand{\HCzeroSE}{0.610}
\newcommand{\HConeSE}{0.625}
\newcommand{\HCtwoSE}{0.864}
\newcommand{\HCthreeSE}{1.266}

\newcommand{\MFVvalue}{881.164}
\newcommand{\MFVeps}{5.733}
\newcommand{\MFVResidual}{1.27e-06}
\newcommand{\MFVAuditResidual}{7.99e-07}

\newcommand{\LogisticM}{882.835}
\newcommand{\LogisticS}{4.101}
\newcommand{\LogisticIdealSE}{1.550}
\newcommand{\LogisticSandSE}{1.704}

\newcommand{\LogisticResidual}{6.94e-15}

\newcommand{\LogisticProdDeltaM}{4.50e-08}
\newcommand{\LogisticProdDeltaS}{6.08e-07}
\newcommand{\InternalLow}{878.434}
\newcommand{\InternalHigh}{878.944}
\newcommand{\PDGLow}{878.203}
\newcommand{\PDGHigh}{879.175}
\newcommand{\HCzeroLow}{878.079}
\newcommand{\HCzeroHigh}{879.299}
\newcommand{\HCthreeLow}{877.423}
\newcommand{\HCthreeHigh}{879.955}
\newcommand{\LogisticSandLow}{881.131}
\newcommand{\LogisticSandHigh}{884.539}
\newcommand{\WBootLow}{878.265}
\newcommand{\WBootHigh}{880.068}

\newcommand{\WHPBLow}{878.224}
\newcommand{\WHPBHigh}{880.121}

\newcommand{\MFVBootZeroDihesionN}{1030}
\newcommand{\MFVHPBZeroDihesionN}{4}
\newcommand{\MFVBootZeroDihesionPct}{5.15}
\newcommand{\MFVHPBZeroDihesionPct}{0.02}

\newcommand{\LogisticInitScale}{4.096}

\newcommand{\MaxLeverage}{0.566}

\newcommand{\TopThreeWeightPct}{78.95}
\newcommand{\EffectiveWeightedN}{2.821}
\newcommand{\BottleCount}{14}
\newcommand{\BeamCount}{6}
\newcommand{\SpaceCount}{1}
\newcommand{\BottleEqualMean}{882.154}
\newcommand{\BeamEqualMean}{889.983}
\newcommand{\SpaceEqualMean}{887.000}
\newcommand{\BottleWeightedMean}{878.517}
\newcommand{\BeamWeightedMean}{888.555}
\newcommand{\SpaceWeightedMean}{887.000}
\newcommand{\BottleWeightPct}{98.28}
\newcommand{\BeamWeightPct}{1.69}
\newcommand{\SpaceWeightPct}{0.03}
\newcommand{\MaxLOOShift}{1.225}
\newcommand{\MaxLOOStudy}{Gonzalez 2021}
\newcommand{\Nboot}{20000}

\newcommand{\MFVBootValidN}{20000}

\newcommand{\MFVBootMCSELow}{0.045}
\newcommand{\MFVBootMCSEHigh}{0.025}
\newcommand{\MFVBootLow}{878.911}
\newcommand{\MFVBootHigh}{883.428}

\newcommand{\MFVHPBValidN}{20000}

\newcommand{\MFVHPBMCSELow}{0.020}
\newcommand{\MFVHPBMCSEHigh}{0.030}
\newcommand{\MFVHPBLow}{879.714}
\newcommand{\MFVHPBHigh}{884.033}

\newcommand{\LogBootLow}{881.198}
\newcommand{\LogBootHigh}{884.679}

\newcommand{\LogHPBLow}{881.161}
\newcommand{\LogHPBHigh}{886.759}

\begin{document}
	
\setstretch{1.0}

\title[Information-loss location and curvature-scale analysis]{Information-Loss Location Estimation and Curvature-Scale Analysis of Physical Measurements: A Gaussian, Cauchy, and Logistic Neutron-Lifetime Benchmark}

\author*[1]{\fnm{Victor V.} \sur{Golovko}\,\href{https://orcid.org/0000-0003-4605-7937}{\textsc{ORCID}: 0000-0003-4605-7937}}\email{victor.golovko@cnl.ca}

\affil*[1]{\orgname{Canadian Nuclear Laboratories}, \orgaddress{\city{Chalk River}, \state{Ontario}, \country{Canada}}}

\abstract{We develop and implement a two-step location--scale analysis for repeated
	physical measurements when the underlying distribution is not known. For a
	chosen substitute distribution, stationarity of population Kullback--Leibler
	information loss at fixed scale motivates the location score, which we apply
	to the observed measurements through empirical cross-entropy. Steiner's
	curvature rule then defines a companion scale separately. We place Gaussian,
	Cauchy, and logistic substitutes in one common construction and implement
	them reproducibly on a physics benchmark. For the logistic substitute, we
	pair the established bounded location score with a scale defined by the
	curvature rule rather than by a fitted tuning constant or a scale-likelihood
	equation. For Gaussian measurements with different quoted uncertainties, an
	experiment-level curvature convention and a stated variance mapping give an
	exact algebraic connection to a standard residual-based variance estimate for a weighted fit.
	Applied to a published 21-measurement neutron-lifetime compilation,
	the inverse-variance weighted, most frequent value, and logistic
	locations are 
	\WeightedMean, \MFVvalue, and \LogisticM~s, respectively. These values are
	benchmark results of the compared constructions and are not proposed as a
	new recommended neutron lifetime.}


\maketitle

\section{Introduction}
Physicists often need one representative value from several measurements of the same quantity. Suppose experiment $i$ reports a value $x_i$ with standard uncertainty $\sigma_i$. If the measurements are independent and the quoted uncertainties describe Gaussian errors well, inverse-variance weighting gives more weight to the more precise measurements. The Particle Data Group (PDG) uses this principle and enlarges the combined uncertainty when the measurements scatter more than expected \cite{PDG2024}. The neutron dataset used below follows the 21-point compilation analyzed by Zhang et al. from the 2022-era literature \cite{Workman2022,Zhang2022}; it is not intended to reproduce a later PDG selection. Combining measurements becomes harder when the error distribution is not Gaussian, when a few observations lie far from the main cluster, or when the quoted uncertainties do not capture the full experiment-to-experiment variation.

Robust statistics offers another approach: it reduces the influence of unusual observations instead of deleting them \cite{Huber1964,RousseeuwVerboven2002}. Steiner developed the most frequent value (MFV) as a robust estimate of
location; in his later treatment, the associated scale is termed the
dihesion \cite{Steiner1973,Steiner1988}. Earlier work derived the MFV location equation from a KL information-loss argument with a Cauchy substitute, retained Steiner's curvature rule for the associated scale, and combined MFV with bootstrap uncertainty estimation \cite{Golovko2025Biomolecules}. HPB was subsequently introduced as a resampling procedure that also perturbs each selected measurement according to its quoted uncertainty \cite{Golovko2025HPB}. Robust statistics uses the term \emph{M-estimation} for methods that estimate a location or other parameter by solving equations built from the measurement residuals. In the logistic form used here, those residual contributions are transformed by a smooth bounded score. Logistic M-estimation and related smooth scale constructions are established methods \cite{RousseeuwVerboven2002}. These ingredients are therefore prior work rather than novelty claims of the present article.

The article-specific contribution is to place three analytical substitutes in one explicit two-step construction and to show what that construction yields in each case. The Gaussian cases provide conventional controls and, under an adopted experiment-level curvature convention plus a stated variance mapping, an exact algebraic connection to a standard residual-based variance estimate for a weighted fit. This estimate is conventionally called HC0 and is defined in the Methods section. The Cauchy case serves as the established MFV reference. For the logistic substitute, we retain the established bounded $\tanh$ location score but define its companion scale by Steiner's curvature condition, which yields the coupled equation $\langle\tanh^2[(x-M)/(2S)]\rangle=1/3$. Thus the logistic contribution is not a new logistic score or a new general class of M-estimators; it is the embedding of that score in the common KL-location/curvature-scale construction, together with a prescribed and reproducible implementation. The factor $1/3$ is fixed by the curvature calculation rather than selected as a robustness tuning constant.

The neutron lifetime provides a controlled physics benchmark because beam, bottle, and space-based measurements rely on different experimental methods and have shown long-standing tension \cite{WietfeldtGreene2011,RajanDesai2020}. Zhang et al. assembled the 21-measurement compilation used here and reported an MFV result that provides a direct implementation check \cite{Zhang2022}. We apply the three constructions to this fixed dataset to compare their numerical behavior and interpretation; we do not propose a new recommended neutron lifetime or a general validation study of estimator bias, coverage, or robustness. The Gaussian estimator with quoted uncertainties gives more weight to measurements reported with smaller uncertainties, whereas the deterministic Cauchy/MFV and logistic estimators give equal status to the reported central values. If the measurements represent physically distinct populations, a single pooled location is not an adequate description.

The rest of the article is organized as follows.
Section~\ref{sec:methods} presents the analysis framework and its
implementation.
Section~\ref{sec:data} introduces the 21-measurement neutron-lifetime
benchmark, and Section~\ref{sec:results} reports the numerical results.
Section~\ref{sec:discussion} discusses their interpretation and limitations,
while Section~\ref{sec:conclusions} summarizes the main findings.
Appendices~\ref{app:uncertainty}--\ref{app:numerical} provide supporting
technical details.

\section{Methods}
\label{sec:methods}

This section develops the common analysis framework used throughout the
article. We first define the information-loss construction and separate the
location equation from the curvature-based scale definition. We then apply
the same procedure to Gaussian, Cauchy/MFV, and logistic substitutes, showing
how the three cases differ in their weighting, score functions, and fitted
scales. Finally, we define the uncertainty estimates, resampling procedures,
and numerical rules used to obtain the results reported below.

\subsection{Information loss and the general location--scale construction}
Suppose repeated measurements of a physical quantity follow an unknown probability distribution $f(x)$. A Gaussian distribution often provides a useful first approximation, especially when many small independent effects contribute to the measurement error, but the true distribution need not be Gaussian. We therefore introduce an analytical substitute $g(x;M,S)$, where $M$ is a location parameter and $S>0$ is a scale parameter.

The construction used here has two distinct steps. First, for a fixed scale $S$, we derive the location estimating equation from stationarity of the KL information loss with respect to $M$. A stationary point need not be a global minimum unless additional curvature or convexity conditions hold; those conditions are checked where they are used below. Second, after the location equation has been obtained, we use Steiner's curvature condition to define a companion scale. The second step is not, in general, equivalent to minimizing KL divergence with respect to $S$. Keeping these two operations separate is important for the interpretation of the framework.

Operationally, every example follows the same sequence: choose a substitute density, derive its location score, convert the population score balance to a finite-sample estimating equation, apply the curvature rule to define a companion scale, solve the resulting equations with a stated numerical convention, and construct uncertainty separately. This ordering matters because the scale parameter and the uncertainty of the estimated location are different objects. Steiner introduced the term \emph{dihesion} for the scale accompanying the Cauchy/MFV construction. We retain that historical use for $\varepsilon$ and, explicitly as a terminology extension of this manuscript, use \emph{Gaussian dihesion} and \emph{logistic dihesion} for the analogous curvature-defined scales of the other substitutes. This extension does not imply that those names are established terminology outside the MFV context. In every case, a dihesion is a fitted scale, not the quoted uncertainty of an individual measurement or the standard error of the fitted location.

Kullback and Leibler introduced a measure of the information lost when one probability distribution is used in place of another \cite{KullbackLeibler1951}. We use
\begin{equation}
D_{\mathrm{KL}}(f\Vert g)
=
\int_{-\infty}^{\infty}
f(x)\ln\!\left[\frac{f(x)}{g(x;M,S)}\right]dx .
\label{eq:KL}
\end{equation}
For fixed $S$, a smaller $D_{\mathrm{KL}}$ means that the substitute represents $f$ more closely in the KL sense as we vary $M$. Separating the logarithm gives
\[
\begin{aligned}
	D_{\mathrm{KL}}
	&=
	\int f(x)\ln f(x)\,dx \\
	&\quad
	-
	\int f(x)\ln g(x;M,S)\,dx .
\end{aligned}
\]
The first integral does not depend on $M$. The location therefore follows from the parameter-dependent second term.

To interpret the population derivative below literally, we assume that the cross-entropy is finite in a neighborhood of the location under consideration, that the required derivatives of $\ln g(x;M,S)$ exist, and that differentiation can be interchanged with integration, for example through an $f$-integrable dominating function. Where a second derivative is used, the analogous integrability and dominated-differentiation condition is required for that derivative. If these population regularity conditions are not available, we use the KL argument only as formal motivation; the empirical cross-entropy in Equation~\eqref{eq:empirical_cross_entropy} and its finite-sample derivatives define the calculations actually implemented.

At a stationary location,
\[
\frac{\partial D_{\mathrm{KL}}}{\partial M}=0.
\]
Differentiating gives
\[
\frac{\partial D_{\mathrm{KL}}}{\partial M}
=
-\int f(x)\frac{\partial}{\partial M}\ln g(x;M,S)\,dx .
\]
Because
\[
\frac{\partial}{\partial M}\ln g
=
\frac{1}{g}\frac{\partial g}{\partial M}
=
\frac{g_M}{g},
\]
the general location condition is
\begin{equation}
\int_{-\infty}^{\infty}
\frac{g_M}{g}f(x)\,dx
=0 .
\label{eq:location_general}
\end{equation}
The ratio $g_M/g$ is the \emph{location score}. It describes how an observation at $x$ contributes to the location equation. Different substitutes give different score functions and therefore different location estimators. At fixed positive scale, the Gaussian score grows linearly with the residual, the logistic score is bounded, and the Cauchy score redescends toward zero in the far tails.

A stationary point alone does not establish a minimum in $M$. Differentiating again gives
\[
\frac{\partial^2 D_{\mathrm{KL}}}{\partial M^2}
=
\int\left(\frac{g_M}{g}\right)^2 f(x)\,dx
-
\int\frac{g_{MM}}{g}f(x)\,dx .
\]
The first term is nonnegative. Steiner used the second curvature contribution to define a scale associated with the location estimator by imposing
\begin{equation}
\int_{-\infty}^{\infty}
\frac{g_{MM}}{g}f(x)\,dx
=0.
\label{eq:steiner_scale}
\end{equation}
We use Equation~\eqref{eq:steiner_scale} as a \emph{separate scale-defining rule} \cite{Steiner1988,Golovko2025Biomolecules}. It is a curvature construction in the location direction, not the scale-stationarity equation $\partial D_{\mathrm{KL}}/\partial S=0$. The two rules happen to give the same scale in some special cases, including the ordinary Gaussian example below, but they differ in general.

The curvature balance has a simple operational interpretation for the Gaussian, Cauchy, and logistic substitutes used below. For a fixed trial location, observations sufficiently close to the center and observations farther away contribute with opposite signs to $g_{MM}/g$. Changing the scale changes their standardized distances and therefore changes that balance. The curvature rule selects the scale at which the empirical positive and negative curvature contributions cancel. The resulting scale is thus the width required by the chosen substitute to balance the observed residual pattern around the fitted location. For MFV, $\varepsilon$ sets the residual scale over which the Cauchy score moves from its central response toward tail attenuation and eventual redescending behavior. For the logistic construction, $S$ sets the transition from the approximately linear central part of the $\tanh$ score to its bounded tail plateau. These are operational spread parameters of the fitted location--scale construction, not uncertainties of the estimated location.

Although the derivatives in Equation~\eqref{eq:steiner_scale} are taken with respect to $M$, the ratio $g_{MM}/g$ also depends on $S$. The location equation and curvature equation therefore form a coupled estimating system for $(M,S)$. This language is useful in practice: $M$ is KL-motivated through location stationarity, whereas $S$ is curvature-defined.

For a finite dataset, the population expectations in Equations~\eqref{eq:location_general} and \eqref{eq:steiner_scale} are unavailable because $f$ is unknown. We replace them by empirical averages. Formally, define the empirical probability measure
\[
F_n=\frac{1}{n}\sum_{i=1}^{n}\delta_{x_i},
\]
where $\delta_{x_i}$ denotes a unit point mass at $x_i$. For any function $\varphi$,
\[
\int \varphi(x)\,dF_n(x)
=
\frac{1}{n}\sum_{i=1}^{n}\varphi(x_i).
\]
This identity converts the population estimating equations into finite sums. We use the empirical measure only for this conversion; we do not evaluate a KL divergence between a discrete empirical measure and a continuous density.

Equivalently, for a fixed positive scale the computable finite-sample location objective is the empirical cross-entropy
\begin{equation}
Q_n(M;S)
=
-\frac{1}{n}\sum_{i=1}^{n}\ln g(x_i;M,S).
\label{eq:empirical_cross_entropy}
\end{equation}
Its location derivative is
\[
\frac{\partial Q_n}{\partial M}
=
-\frac{1}{n}\sum_{i=1}^{n}
\frac{g_M(x_i;M,S)}{g(x_i;M,S)}.
\]
Thus the finite-sample score equation used below is exactly the stationarity condition for Equation~\eqref{eq:empirical_cross_entropy}. The population KL argument motivates the expected score, while Equation~\eqref{eq:empirical_cross_entropy} is the quantity that connects that argument to the observed sample.

Table~\ref{tab:workflow} summarizes the implementable workflow before the detailed derivations and separates fitted scale parameters from uncertainty summaries of the estimated location.

\begin{table*}[t]
	\centering
	\begin{singlespace}
		
		\caption{Summary of the four constructions used in the physics benchmark.
			MFV is the established Cauchy reference. For the logistic case, the location
			score is established; its pairing here with the curvature-defined scale is
			the article-specific extension.}
		\label{tab:workflow}
		
		\small
		\renewcommand{\arraystretch}{1.08}
		
		\begin{tabularx}{\textwidth}{
				@{}
				L{2.0cm}
				L{2.2cm}
				L{1.8cm}
				L{2.3cm}
				Y
				@{}
			}
			\toprule
			Construction &
			Input and location score &
			Companion scale &
			Numerical solution &
			Uncertainty summary \\
			\midrule
			
			Equal-status Gaussian &
			$x_i$; linear score &
			empirical RMS scale &
			closed form &
			control spread only; not a location SE \\
			
			Quoted-uncertainty Gaussian &
			$x_i,\sigma_i$; inverse-variance-weighted linear score &
			unitless curvature factor $S_G$ &
			closed form &
			internal, PDG/Birge, and HC0--HC3 estimates \\
			
			Cauchy/MFV &
			$x_i$; redescending Cauchy score &
			Steiner dihesion $\varepsilon$ &
			fixed-point iteration &
			central 68.27\% descriptive bootstrap and HPB percentile ranges \\
			
			Logistic &
			$x_i$; bounded $\tanh$ score &
			$S_{\log}$ with $\langle\tanh^2 z\rangle=1/3$ &
			damped-Newton iteration &
			sandwich SE and descriptive resampling ranges \\
			
			\bottomrule
		\end{tabularx}
		
	\end{singlespace}
\end{table*}

\subsection{Gaussian substitute with equal measurement status}
We first test the construction with an ordinary Gaussian substitute. This case provides a useful check because it should recover familiar statistics. We use
\[
g(x;M,S)
=
\frac{1}{\sqrt{2\pi}S}
\exp\!\left[-\frac{(x-M)^2}{2S^2}\right],
\]
where $M$ is the center and $S>0$ is the width.

We obtain the location score most easily by differentiating $\ln g$:
\[
\ln g
=
-\frac12\ln(2\pi)-\ln S-\frac{(x-M)^2}{2S^2}.
\]
Only the last term depends on $M$, so
\[
\frac{g_M}{g}
=
\frac{\partial}{\partial M}\ln g
=
\frac{x-M}{S^2}.
\]
Substituting this expression into Equation~\eqref{eq:location_general} gives
\[
\int (x-M)f(x)\,dx=0.
\]
Since $\int f(x)\,dx=1$, the solution is
\[
M=\int x f(x)\,dx=E_f[X].
\]
For a finite sample, we replace the expectation by the sample average and recover the arithmetic mean,
\[
M=\frac{1}{n}\sum_{i=1}^{n}x_i.
\]

We now use Equation~\eqref{eq:steiner_scale} to determine the Gaussian scale. The identity
\[
\frac{g_{MM}}{g}
=
\left(\frac{g_M}{g}\right)^2
+
\frac{\partial^2}{\partial M^2}\ln g
\]
gives
\[
\frac{g_{MM}}{g}
=
\frac{(x-M)^2}{S^4}-\frac{1}{S^2}.
\]
Substituting this result into Equation~\eqref{eq:steiner_scale} and multiplying by $S^4$ gives
\[
\int\left[(x-M)^2-S^2\right]f(x)\,dx=0.
\]
Therefore
\[
S^2
=
\int (x-M)^2f(x)\,dx.
\]
For a finite sample,
\[
S
=
\sqrt{\frac{1}{n}\sum_{i=1}^{n}(x_i-M)^2}.
\]
The denominator is $n$, not $n-1$, because Equation~\eqref{eq:steiner_scale} gives this expression directly. The result is the root-mean-square (RMS) residual of the empirical distribution, not the usual unbiased estimate of the sample standard deviation. The ordinary Gaussian substitute therefore gives the familiar arithmetic mean and a Gaussian RMS scale, as expected. We use this equal-status Gaussian case only as an algebraic reference and do not assign it a separate location standard error in the neutron-lifetime comparison.

\subsection{Gaussian substitute with quoted experimental uncertainties}
Physical measurements rarely have equal precision. Suppose $n$ independent experiments measure the same physical quantity. Experiment $i$ reports a value $x_i$ with standard uncertainty $\sigma_i$. Because all experiments target the same physical quantity, we fit one common location $M$ while retaining the relative precision information carried by the individual $\sigma_i$.

To keep the integration variables separate from the observed values, let $y_i$ denote a possible outcome of experiment $i$. We use the Gaussian substitute
\[
g_i(y_i;M,S_G)
=
\frac{1}{\sqrt{2\pi}\,S_G\sigma_i}
\exp\!\left[-\frac{(y_i-M)^2}{2S_G^2\sigma_i^2}\right],
\]
where $S_G>0$ is dimensionless. In the analytical substitute, $S_G$ multiplies every quoted uncertainty and therefore preserves their relative sizes. This parameterization is a convenient way to derive the location and curvature equations. The fitted curvature quantity obtained below should not, however, be interpreted automatically as a calibrated physical estimate of a common multiplier of the experimental uncertainties.

Let $f_i(y_i)$ denote the unknown repeated-measurement distribution for experiment $i$. Under independence, define the joint true and substitute densities as
\[
\begin{aligned}
f_{\mathrm{joint}}(\mathbf y)
&=\prod_{i=1}^{n}f_i(y_i),\\
g_{\mathrm{joint}}(\mathbf y;M,S_G)
&=\prod_{i=1}^{n}g_i(y_i;M,S_G).
\end{aligned}
\]
where $\mathbf y=(y_1,\ldots,y_n)$. The logarithm of a product is a sum, so the joint KL divergence separates into individual contributions:
\[
\begin{aligned}
D_{\mathrm{KL}}(f_{\mathrm{joint}}\Vert g_{\mathrm{joint}})
&=
\sum_{i=1}^{n}\int f_i(y_i)\\
&\quad\times
\ln\!\left[\frac{f_i(y_i)}
{g_i(y_i;M,S_G)}\right]dy_i .
\end{aligned}
\]
This factorization shows why independent measurements add their KL contributions.

We first determine the common location while holding $S_G$ fixed. For one Gaussian,
\[
\frac{\partial}{\partial M}\ln g_i
=
\frac{y_i-M}{S_G^2\sigma_i^2}.
\]
The population location condition is therefore
\[
\sum_{i=1}^{n}
\int
\frac{y_i-M}{S_G^2\sigma_i^2}
f_i(y_i)\,dy_i
=0.
\]
For the observed vector $\mathbf x=(x_1,\ldots,x_n)$, the empirical estimating equation becomes
\[
\sum_{i=1}^{n}
\frac{x_i-M}{S_G^2\sigma_i^2}
=0.
\]
The common factor $1/S_G^2$ cancels. Defining the inverse-variance weight $w_i=1/\sigma_i^2$ gives
\[
\sum_iw_ix_i-M\sum_iw_i=0,
\]
and therefore
\begin{equation}
M_w
=
\frac{\sum_iw_ix_i}{\sum_iw_i},
\qquad
w_i=\frac{1}{\sigma_i^2}.
\label{eq:weightedmean}
\end{equation}
Thus KL stationarity in the location direction gives the familiar inverse-variance weighted mean. A smaller quoted uncertainty produces a larger weight and therefore gives that measurement greater potential influence on the fitted common value. The regression term \emph{leverage} is introduced below for a quantitative measure of this weight-based influence.

We next apply Steiner's scale rule to the additive KL contributions. Their curvature in $M$ is
\[
\begin{aligned}
\frac{\partial^2D_{\mathrm{KL}}}{\partial M^2}
&=
\sum_i\int
\left(\frac{g_{i,M}}{g_i}\right)^2
f_i(y_i)\,dy_i\\
&\quad-
\sum_i\int
\frac{g_{i,MM}}{g_i}
f_i(y_i)\,dy_i .
\end{aligned}
\]
The extension used here sets the summed second curvature contribution to zero,
\[
\sum_i
\int\frac{g_{i,MM}}{g_i}f_i(y_i)\,dy_i
=0.
\]
This is a curvature rule applied to the sum of the individual KL contributions. It is not the same as either setting $G_{MM}/G=0$ for the full product density or solving $\partial D_{\mathrm{KL}}/\partial S_G=0$.

For the $i$th Gaussian,
\[
\frac{g_{i,MM}}{g_i}
=
\frac{(y_i-M)^2}{S_G^4\sigma_i^4}
-
\frac{1}{S_G^2\sigma_i^2}.
\]
Evaluating the empirical equation at $M=M_w$ gives
\[
\sum_i
\left[
\frac{(x_i-M_w)^2}{S_G^4\sigma_i^4}
-
\frac{1}{S_G^2\sigma_i^2}
\right]
=0.
\]
Solving for $S_G^2$ yields
\[
S_G^2
=
\frac{\displaystyle\sum_i (x_i-M_w)^2/\sigma_i^4}
{\displaystyle\sum_i1/\sigma_i^2}.
\]
With residuals $e_i=x_i-M_w$ and the weights from Equation~\eqref{eq:weightedmean},
\begin{equation}
S_G^2
=
\frac{\sum_iw_i^2e_i^2}{\sum_iw_i}.
\label{eq:steiner_gauss_scale}
\end{equation}
As an explicit extension of Steiner's MFV terminology, we refer to $S_G$ as the \emph{Gaussian dihesion factor} to keep the notation parallel to the Cauchy and logistic cases. It is dimensionless, but Equation~\eqref{eq:steiner_gauss_scale} defines it directly from the fitted residuals through the curvature rule; it is not an unbiased estimate of a physical common uncertainty multiplier.

This distinction can be seen directly under a literal common-scale Gaussian model. Let $\tau>0$ denote the hypothetical common multiplier of the quoted experimental uncertainties, and suppose
\[
x_i=M+\eta_i,
\qquad
E(\eta_i)=0,
\qquad
\operatorname{Var}(\eta_i)=\tau^2\sigma_i^2,
\]
and define
\[
W=\sum_iw_i,
\qquad
h_i=\frac{w_i}{W}.
\]
For the weighted-mean residual $e_i=x_i-M_w$,
\[
E(e_i^2)
=
\tau^2\left(\frac{1}{w_i}-\frac{1}{W}\right).
\]
Substitution into Equation~\eqref{eq:steiner_gauss_scale} gives
\begin{equation}
E(S_G^2)
=
\tau^2\left(1-\sum_i h_i^2\right).
\label{eq:sg_expectation}
\end{equation}
The factor in parentheses is smaller than one whenever the weights are concentrated. In the limiting case in which one observation carries essentially all of the weight, it approaches zero. Equation~\eqref{eq:sg_expectation} therefore prevents us from interpreting the fitted $S_G$ as a calibrated estimate of $\tau$ without an additional correction and validation.

The useful result of Equation~\eqref{eq:steiner_gauss_scale} is instead an algebraic relation to a familiar covariance diagnostic for the weighted location. The usual internal variance of $M_w$ is $1/W$. If, as an additional mapping, we define
\[
V_G(M_w)=\frac{S_G^2}{W},
\]
then
\begin{equation}
V_G(M_w)
=
\frac{\sum_iw_i^2e_i^2}{W^2}.
\label{eq:hc0identity}
\end{equation}
Under the adopted experiment-level additive curvature convention and the additional mapping $V_G=S_G^2/W$, Equation~\eqref{eq:hc0identity} is algebraically identical to a standard residual-based variance estimate for a weighted fit \cite{White1980}. In statistics this estimate is called HC0, the original heteroscedasticity-consistent or \emph{sandwich} variance. Here ``heteroscedasticity-consistent'' means that the variance calculation does not require the residual scatter to be the same for every measurement, and ``intercept-only'' means that the weighted least-squares model fits only one common constant rather than a slope or other predictors. HC0 is an established covariance estimator; we do not claim it as a new method. The connection is therefore an exact algebraic identity under these stated choices, not a derivation of HC0 as a uniquely curvature-calibrated uncertainty. It does not by itself show that HC0 has correct uncertainty coverage for a sample of this size, provide a physical common-width model, or establish a general theory of residual-based covariance estimates.

\subsection{Relation to the PDG/Birge scale factor and highly weighted measurements}
The uncertainty constructions compared here answer different practical questions while leaving the inverse-variance weighted location unchanged. The \emph{internal} standard error $1/\sqrt{W}$ asks how precisely the weighted mean would be known if the quoted $\sigma_i$ values were complete, independent Gaussian standard uncertainties. The PDG/Birge prescription keeps that same weighted mean but applies one global inflation factor when the observed scatter is larger than those quoted uncertainties predict. HC0--HC3 instead estimate the variance of the fitted weighted location from the observed residuals. HC0 uses the raw residual contributions. HC1 multiplies HC0 by the finite-sample factor $n/(n-1)$ because one common location has been estimated from the same $n$ measurements. HC2 and HC3 progressively increase the contribution from measurements that carry a large fraction of the total inverse-variance weight. In regression terminology that fraction is called \emph{leverage}; in this one-parameter weighted fit it is simply $h_i=w_i/W$. Thus internal, PDG/Birge, and HC0--HC3 are not competing location estimators; they are different uncertainty constructions around the same weighted location. The complete HC formulas are collected in Appendix~\ref{app:uncertainty}.

The PDG/Birge prescription starts from the same weighted mean in Equation~\eqref{eq:weightedmean}, while each measurement keeps its own $\sigma_i$. It measures the scatter around the weighted mean in units of the reported uncertainties with
\begin{equation}
\chi^2
=
\sum_i\frac{(x_i-M_w)^2}{\sigma_i^2}
=
\sum_iw_ie_i^2.
\label{eq:chi2}
\end{equation}
Each term compares one residual with the uncertainty reported for that measurement. When the total scatter exceeds the Gaussian expectation, the PDG prescription enlarges the internal uncertainty of the weighted mean by
\begin{equation}
S_{\mathrm{PDG}}
=
\max\!\left(1,\sqrt{\frac{\chi^2}{n-1}}\right),
\qquad
u_{\mathrm{PDG}}
=
\frac{S_{\mathrm{PDG}}}{\sqrt{W}}.
\label{eq:pdgscale}
\end{equation}
The denominator $n-1$ reflects the loss of one degree of freedom when we estimate the common location from the same $n$ measurements. The lower bound $S_{\mathrm{PDG}}\ge1$ prevents this prescription from shrinking the quoted combined uncertainty when the observed scatter is smaller than expected. Particle-data evaluations use this scale-factor approach routinely \cite{PDG2024,RajanDesai2020}.

The Gaussian curvature quantity $S_G$ and the PDG/Birge factor both use the same unequal-uncertainty dataset, but they summarize residuals differently and have different interpretations. Equation~\eqref{eq:chi2} contains $w_ie_i^2$, whereas the numerator of Equation~\eqref{eq:steiner_gauss_scale} contains $w_i^2e_i^2$. The extra factor $w_i$ makes the curvature quantity more sensitive to disagreement from measurements with small quoted uncertainties. Their numerical values can therefore differ even though they use the same $x_i$, the same $\sigma_i$, and the same weighted location $M_w$; Equation~\eqref{eq:sg_expectation} also shows why $S_G$ should not be read as a calibrated counterpart of $S_{\mathrm{PDG}}$.

We use an equal-uncertainty calculation only as an algebraic check; the neutron analysis itself keeps the reported unequal uncertainties. If all $\sigma_i$ are set equal, all weights become equal and Equation~\eqref{eq:steiner_gauss_scale} reduces to
\[
S_G^2=\frac{\chi^2}{n}.
\]
For an intercept-only fit, HC1 multiplies the HC0 variance by $n/(n-1)$. The corresponding squared scale is then
\[
S_{\mathrm{HC1}}^2
=
\frac{n}{n-1}S_G^2
=
\frac{\chi^2}{n-1}.
\]
When $\chi^2/(n-1)>1$, the equal-uncertainty limit gives $S_{\mathrm{HC1}}=S_{\mathrm{PDG}}$, or equivalently $S_{\mathrm{HC1}}^2=S_{\mathrm{PDG}}^2=\chi^2/(n-1)$. The PDG prescription additionally requires $S_{\mathrm{PDG}}\ge1$. This is an algebraic limiting connection only. In the actual unequal-uncertainty dataset, Equation~\eqref{eq:steiner_gauss_scale} and the PDG/Birge construction remain distinct.

HC1--HC3 provide finite-sample and high-weight adjustments to the HC0 residual variance~\cite{MacKinnonWhite1985}. In the present weighted fit, $h_i=w_i/W$ is the fraction of the total inverse-variance weight carried by measurement $i$. This quantity is the measurement's leverage: it measures potential influence arising from quoted precision before the size of the residual is considered. The realized leave-one-out shift also depends on the residual:
\[
M_{w,(-i)}-M_w
=
-\frac{h_i}{1-h_i}\,e_i .
\]
Thus a high-precision point need not produce the largest deletion shift unless it also disagrees with the fitted location. HC2 and HC3 increasingly enlarge the residual contribution from measurements with large leverage, that is, measurements carrying a large fraction of the total weight. The complete formulas are collected in Appendix~\ref{app:uncertainty}. These quantities are uncertainty diagnostics for the weighted location, not calibrated common-width multipliers for the individual measurements.

\subsection{Cauchy substitute and the most frequent value}
We next choose a Cauchy substitute,
\[
g(x;M,\varepsilon)
=
\frac{1}{\pi}\frac{\varepsilon}{\varepsilon^2+(x-M)^2},
\]
where $M$ is the location and $\varepsilon>0$ is the scale. We write the scale as $\varepsilon$ to follow the standard MFV notation \cite{Steiner1988,Golovko2025Biomolecules}.

Taking the logarithm gives
\[
\ln g
=
-\ln\pi+\ln\varepsilon
-\ln\!\left[\varepsilon^2+(x-M)^2\right].
\]
Differentiation with respect to $M$ gives
\[
\frac{g_M}{g}
=
\frac{2(x-M)}{\varepsilon^2+(x-M)^2}.
\]
Substituting this score into Equation~\eqref{eq:location_general} and replacing the population expectation by the sample average gives
\[
\sum_i
\frac{x_i-M}{\varepsilon^2+(x_i-M)^2}
=0.
\]
Rearranging the equation isolates $M$:
\[
M
=
\frac{\displaystyle\sum_i
\frac{x_i}{\varepsilon^2+(x_i-M)^2}}
{\displaystyle\sum_i
\frac{1}{\varepsilon^2+(x_i-M)^2}}.
\]
The right-hand side still contains the unknown $M$ and $\varepsilon$, so the equation does not give a one-step solution. We solve it by fixed-point iteration,
\begin{equation}
M^{(k+1)}
=
\frac{\displaystyle\sum_i
\frac{x_i}{(\varepsilon^{(k)})^2+(x_i-M^{(k)})^2}}
{\displaystyle\sum_i
\frac{1}{(\varepsilon^{(k)})^2+(x_i-M^{(k)})^2}}.
\label{eq:mfvMiter}
\end{equation}

The scale follows from Steiner's curvature condition. For the Cauchy substitute,
\[
\frac{g_{MM}}{g}
=
\frac{2\left[3(x-M)^2-\varepsilon^2\right]}
{\left[\varepsilon^2+(x-M)^2\right]^2}.
\]
The common factor 2 cancels in Equation~\eqref{eq:steiner_scale}, so the sample scale equation is
\[
\sum_i
\frac{3(x_i-M)^2-\varepsilon^2}
{\left[\varepsilon^2+(x_i-M)^2\right]^2}
=0.
\]
Solving this equation for $\varepsilon^2$ gives
\[
\varepsilon^2
=
3\,
\frac{\displaystyle\sum_i
\frac{(x_i-M)^2}
{\left[\varepsilon^2+(x_i-M)^2\right]^2}}
{\displaystyle\sum_i
\frac{1}
{\left[\varepsilon^2+(x_i-M)^2\right]^2}}.
\]
We therefore update the scale with
\begin{equation}
(\varepsilon^{(k+1)})^2
=
3\,
\frac{\displaystyle\sum_i
\frac{(x_i-M^{(k)})^2}
{[(\varepsilon^{(k)})^2+(x_i-M^{(k)})^2]^2}}
{\displaystyle\sum_i
\frac{1}
{[(\varepsilon^{(k)})^2+(x_i-M^{(k)})^2]^2}}.
\label{eq:mfvSiter}
\end{equation}
For the numerical implementation used here, we follow the published single-path MFV iteration and start the location at the sample median,
\[
M^{(0)}=\operatorname{median}(x),
\]
with the broad positive scale
\[
\varepsilon^{(0)}
=
\frac{\sqrt{3}}{2}(x_{\max}-x_{\min}).
\]
At each iteration, Equations~\eqref{eq:mfvMiter} and \eqref{eq:mfvSiter} are both evaluated from the same current pair $(M^{(k)},\varepsilon^{(k)})$. The iteration stops when
\[
\begin{aligned}
|M^{(k+1)}-M^{(k)}|&<10^{-5}\ {\rm s},\\
|\varepsilon^{(k+1)}-\varepsilon^{(k)}|&<10^{-5}\ {\rm s}.
\end{aligned}
\]
The threshold is far below the precision at which the fitted quantities are reported and many orders of magnitude below the experimental uncertainty scale of the input data. Further numerical refinement therefore does not affect the scientifically relevant reported values. We use the same initialization, iteration, and stopping criterion for the original dataset and for every resampled dataset.

Some resamples contract toward zero dihesion. If the iteration satisfies the stopping criterion with a raw scale $\varepsilon\le10^{-5}$~s, we report $\varepsilon=0$ and treat this as the zero-scale limiting case. For positive-scale solutions, bounded equation residuals are evaluated at the stopping point as numerical diagnostics; they do not define a second solution or alter the reported $M$ or $\varepsilon$. The initialization and iterative scheme are part of the numerical definition of the estimator, and we report the pair reached by this prescribed iteration rather than attempting to enumerate alternative stationary roots. Appendix~\ref{app:numerical} gives the residual diagnostics, iteration cap, degenerate-sample handling, and unresolved terminal states.

The name \emph{most frequent value} refers to this estimator and should not be read as the empirical mode or the most repeated observed number. The MFV is defined by the coupled location--scale equations together with the stated initialization, iterative update rules, and stopping criterion, so it can lie between observed data values even when no observation is repeated.

Equation~\eqref{eq:mfvMiter} has the form of a weighted mean, but its residual-dependent weights change with the current $M$ and $\varepsilon$. For fixed positive $\varepsilon$ and large $|x_i-M|$, a point's weight falls approximately as the inverse square of its distance from $M$, and the Cauchy location-score contribution tends back toward zero. Robust statistics calls this fixed-scale score behavior \emph{redescending}. Because $\varepsilon$ is fitted jointly with $M$, this fixed-scale property
alone does not tell us how the full joint estimator responds to arbitrarily
distant observations or how robust it remains when outliers are present.

\subsection{Logistic substitute: established score and curvature-defined scale}
The logistic substitute provides a smooth intermediate case between Gaussian and Cauchy behavior. In robust statistics, M-estimation means estimating parameters by solving equations built from residual-dependent score functions. The logistic form uses a smooth bounded score, and related logistic scale constructions are established \cite{RousseeuwVerboven2002}. We therefore do not claim the logistic score itself as new. The extension studied here is narrower: we obtain the location equation from the same fixed-scale KL stationarity used for the other substitutes and then define the companion logistic scale with Steiner's curvature condition rather than with a separately tuned M-scale or scale-likelihood equation. Its tails are heavier than Gaussian tails, and its location score remains bounded, but the score does not return to zero for very distant observations. We use
\[
g(x;M,S)
=
\frac{1}{4S}\sech^2\!\left(\frac{x-M}{2S}\right).
\]
Define
\[
z=\frac{x-M}{2S}.
\]
Since $d\ln(\sech z)/dz=-\tanh z$ and $\partial z/\partial M=-1/(2S)$,
\[
\frac{g_M}{g}
=
\frac{1}{S}\tanh z.
\]
The finite-sample location equation is therefore
\begin{equation}
\Psi_1(M,S)
=
\frac{1}{n}\sum_i
\tanh\!\left(\frac{x_i-M}{2S}\right)
=0.
\label{eq:loglocation}
\end{equation}
For a large positive or negative residual, $\tanh z$ approaches $+1$ or $-1$. A distant point therefore cannot make an arbitrarily large contribution to the location equation. Unlike the Cauchy score, however, the logistic score does not return toward zero in the far tails \cite{RousseeuwVerboven2002}.

Differentiating once more gives
\[
\frac{g_{MM}}{g}
=
\frac{1}{S^2}
\left[\tanh^2z-\frac12\sech^2z\right].
\]
Using $\sech^2z=1-\tanh^2z$, the bracket becomes
\[
\tanh^2z-\frac12(1-\tanh^2z)
=
\frac32\tanh^2z-\frac12.
\]
Steiner's condition therefore requires
\[
\frac32\left\langle\tanh^2z\right\rangle-\frac12=0,
\]
where the angle brackets denote the sample average. Hence
\begin{equation}
\Psi_2(M,S)
=
\frac{1}{n}\sum_i
\tanh^2\!\left(\frac{x_i-M}{2S}\right)
-\frac13
=0.
\label{eq:logscale}
\end{equation}
The value $1/3$ is not a chosen tuning constant; it follows directly from Steiner's curvature equation for the logistic substitute. This distinguishes Equation~\eqref{eq:logscale} from a separately calibrated logistic M-scale. It also differs from ordinary logistic maximum-likelihood scale stationarity: for the same density, differentiating with respect to $S$ gives
\[
\frac{1}{n}\sum_i 2z_i\tanh z_i=1,
\]
rather than Equation~\eqref{eq:logscale}. The study-specific logistic construction is therefore the pairing of the established bounded location score with this curvature-defined companion scale inside the common two-step framework. As an explicit terminology extension of Steiner's MFV usage, we call the fitted $S$ the \emph{logistic dihesion}; this name is not claimed to be established terminology outside the present extension.

Exact two-support samples provide one useful analytical special case. For an exact two-value sample with multiplicities $n_a$ and $n_b$, a finite positive-scale solution exists only when
\begin{equation}
\frac13<\frac{n_a}{n_b}<3.
\label{eq:logistic_two_cluster_domain}
\end{equation}
At the limiting ratios $3{:}1$ and $1{:}3$, the corresponding solution is approached only as $S\to0$, while a more imbalanced exact two-support sample has no finite positive-scale solution. This is an analytically established special case, not a general existence classifier or a model for clustering in the neutron dataset. The present procedure is intended as a prespecified single-location summary. If the data contain physically distinct groups, they should not be forced into one pooled location. Appendix~\ref{app:numerical} gives the short derivation and the corresponding numerical treatment.

For the logistic substitute, the location objective already curves upward for every positive $S$ because
\[
\frac{\partial^2D_{\mathrm{KL}}}{\partial M^2}
=
\frac{1}{2S^2}E_f[\sech^2z]>0.
\]
Thus Equation~\eqref{eq:logscale} defines the logistic scale; we do not need it to make the location a minimum.

We solve Equations~\eqref{eq:loglocation} and \eqref{eq:logscale} with one prescribed damped-Newton iteration. The location starts at the sample median and, when the MAD is positive, the scale starts at $S^{(0)}=\mathrm{MAD}/\ln3$. The iteration stops when
\[
|\Delta M|<10^{-5}~{\rm s}
\qquad\text{and}\qquad
|\Delta S|<10^{-5}~{\rm s}.
\]
As in the MFV calculation, this threshold is far below both the reporting precision and the experimental uncertainty scale of the neutron-lifetime data, so further numerical refinement does not affect the reported result. The equation residuals are evaluated at the stopping point as a numerical diagnostic. The original dataset and every resample use the same damped-Newton algorithm, initialization rule, line search, and stopping criterion. The resulting pair $(M,S)$ is the reported logistic solution. If the prescribed iteration cannot reach the stopping criterion because of a singular or non-finite numerical state, the outcome is classified as numerically unresolved. Appendix~\ref{app:numerical} specifies the Jacobian, tied-sample scale fallback, line search, scale floor, conditioning limit, iteration cap, and terminal classifications.

\subsection{Uncertainty and resampling overview}
Several different quantities appear in the analysis, and they should not be interpreted as interchangeable uncertainties. The reported $\sigma_i$ describes the uncertainty assigned to measurement $i$. A dihesion describes the scale associated with a fitted location equation. A standard error describes how much an estimated location would vary under repeated sampling according to a stated model or a large-sample approximation. A bootstrap percentile range is an empirical range from a specified resampling procedure. Keeping these roles separate is essential when the methods are compared.

The scale symbol $S$ is used locally for the substitute currently under discussion; cross-method comparisons use qualified notation such as $S_G$, $\varepsilon$, and $S_{\log}$. Appendix~\ref{app:uncertainty} summarizes the scale and uncertainty symbols and makes explicit that a curvature-defined dihesion is not automatically a standard error of the fitted location.

For the inverse-variance weighted mean, the nominal internal standard uncertainty is $1/\sqrt{W}$. The PDG/Birge prescription, the curvature-derived HC0 identity, and the HC1--HC3 corrections modify or replace that nominal variance in different ways. They should therefore be described by their specific construction rather than as interchangeable ``68.27\% intervals.''

For the logistic estimator, we can obtain a simple large-sample reference under an exact symmetric logistic model. Write
\[
z=\frac{x-M}{2S},
\qquad
\psi_M(z)=\tanh z.
\]
For an exact logistic distribution, symmetry gives
\[
E(\tanh^2 z)=\frac13,
\qquad
E(\sech^2 z)=\frac23.
\]
The average derivative of the location score is therefore
\[
A_M
=
E\!\left(\frac{\partial\psi_M}{\partial M}\right)
=
-\frac{1}{2S}E(\sech^2 z)
=
-\frac{1}{3S},
\]
while the score variance is
\[
B_M=E(\psi_M^2)=\frac13.
\]
The usual large-sample estimating-equation variance then gives
\[
\operatorname{Var}(\widehat M)
\simeq
\frac{B_M}{nA_M^2}
=
\frac{3S^2}{n},
\]
and hence
\[
\operatorname{SE}_{\mathrm{ideal}}(\widehat M)
=
\frac{\sqrt3\,S}{\sqrt n}.
\]
Under this symmetric logistic model, the location and scale equations have zero first-order cross terms by symmetry. We use this formula only as an exact-logistic large-sample reference. We also estimate the joint covariance of $M$ and $S$ with a standard residual-based covariance calculation known as the \emph{sandwich} form. It combines two ingredients: how much the individual score contributions vary across measurements and how strongly the estimating equations change when the fitted parameters change. Its matrix expression is
\[
\widehat{\mathrm{Cov}}(\widehat{\bm{\theta}})
=
\frac{1}{n}A^{-1}BA^{-T},
\qquad
\bm{\theta}=(M,S)^T.
\]
The sandwich standard error of $M$ is then obtained from the estimated variance of $M$, given by the first diagonal element of this covariance matrix:
\[
\operatorname{SE}_{\rm sand}(M)
=
\sqrt{
	\left[
	\widehat{\mathrm{Cov}}(\widehat{\bm{\theta}})
	\right]_{11}
}.
\] 
In a one-parameter scalar analogy, this is score variability divided by squared score sensitivity: more variable score contributions increase the uncertainty, while a more steeply changing estimating equation decreases it. This is a large-sample, or asymptotic, approximation. Its usual interpretation requires the fitted solution to lie away from a parameter boundary, the sensitivity matrix to be invertible so that small parameter changes can be distinguished, and the study-level contributions to satisfy the independence assumptions of the calculation. Appendix~\ref{app:uncertainty} gives the score vector, the matrices $A$ and $B$, and the complete calculation.

The three location methods therefore summarize the same measurements in different ways and answer different statistical questions. The weighted Gaussian location estimates a common value while giving more influence to measurements with smaller quoted uncertainties. The MFV and logistic locations instead give equal status to the reported central values and do not use the individual $\sigma_i$ in their deterministic location equations. We compare them to show how the result changes with these choices; we do not treat their three numerical locations as interchangeable estimates of one identically defined target quantity.

This distinction also determines method selection. A precision-weighted Gaussian combination is appropriate only when a common physical target is scientifically defensible and the relative quoted uncertainties and covariance assumptions are credible. Equal-status MFV or logistic summaries answer a different question and are appropriate only when giving every reported central value equal status is intentionally part of the analysis. If the measurement methods represent physically distinct populations, a single pooled location is not an appropriate description.

When an equal-status robust location is scientifically appropriate, MFV and logistic encode different prespecified fixed-scale tail responses. The MFV score redescends toward zero, whereas the logistic score remains bounded but approaches a finite nonzero tail contribution. These fixed-scale score properties do not, by themselves, determine how the full jointly fitted location--scale estimators respond to arbitrary contamination. If neither tail response is preferred on physical grounds, we report both as sensitivity summaries rather than selecting the estimator that gives the preferred numerical result.

Ordinary nonparametric bootstrap sampling can be used to study how an estimator changes under case resampling \cite{Efron1979}. For each replicate, we draw $n$ study entries with replacement and recalculate the estimator. If the analysis treats the observed studies as interchangeable draws from a broader population of similar studies, the bootstrap uses the observed compilation as an empirical stand-in for that broader population. This is an assumption of the resampling interpretation and should not be confused with repeated measurement of a fixed set of experiments.

HPB adds a second level of variation \cite{Golovko2025HPB}. After selecting a study with replacement, it perturbs that selected central value according to its quoted uncertainty. For asymmetric uncertainties, if $Z$ is standard normal,
\begin{equation}
\Delta_i
=
\begin{cases}
\sigma_{-,i}Z, & Z<0,\\
\sigma_{+,i}Z, & Z\ge0,
\end{cases}
\qquad
x_i^*=x_i+\Delta_i.
\label{eq:hpb}
\end{equation}
This perturbation is median-centered at the published central value and preserves the quoted lower and upper one-standard-deviation scales. It is not generally mean-centered when $\sigma_{-,i}\ne\sigma_{+,i}$. HPB also assumes that the perturbations are independent and cannot reconstruct unreported correlations or shared systematic effects. We use the resulting distribution as a two-level sensitivity construction: case resampling changes the composition of an empirical study population, while the perturbation step propagates the quoted within-study scales. This is not, by itself, a calibrated sampling distribution for a fixed neutron-lifetime parameter. In particular, if the reported $x_i$ are instead viewed only as already-realized noisy estimates from a fixed set of studies, an additional perturbation layer does not have an automatic repeated-sampling interpretation and can overlap with uncertainty already represented in those estimates. A stronger inferential interpretation would require a model that explicitly represents both within-study and between-study variation, together with any shared correlations; such models are often called hierarchical models.

For both ordinary bootstrap and HPB, each selected row is treated as one complete study record and every replicate is refit with exactly the same deterministic estimator used for the original data. We report the 0.15865 and 0.84135 empirical quantiles only when every replicate is usable; otherwise the corresponding published percentile range is withheld. Appendix~\ref{app:resampling} gives the complete five-step workflow, zero-dihesion and unresolved-fit policy, random-number design, quantile convention, and endpoint Monte Carlo-error calculation. The reported central 68.27\% descriptive HPB percentile ranges summarize the resampling distributions. They are not presented as confidence intervals because their repeated-sampling coverage has not been calibrated.

\section{Neutron-lifetime dataset}
\label{sec:data}
We use the 21 neutron-lifetime measurements listed in Table~1 of Zhang et al. \cite{Zhang2022}. Their compilation draws on the 2022 Particle Data Group evaluation and the earlier meta-analysis by Rajan and Desai \cite{Workman2022,RajanDesai2020}. Table~\ref{tab:data} reproduces the central values, quoted uncertainty entries, and method labels used in our calculation. Where a journal source is available, the study column also cites the primary measurement paper so that readers can trace each entry to the original experiment. For this controlled worked example, we use the uncertainty components as tabulated rather than reconstructing experiment-level covariance matrices or reclassifying every component as statistical or systematic. For inverse-variance weighting, multiple quoted components are combined in quadrature, which treats those components as independent standard-uncertainty scales. For asymmetric entries, we first average the quoted lower and upper asymmetric component for the symmetric weighted-mean calculation; HPB instead keeps the lower and upper sides separate. Across studies, the calculation assumes independence because shared covariance information is not available in the compilation. We do not infer from the summary table which quoted components are statistical, experiment-specific systematic, or shared systematic contributions. The HPB calculation should therefore be read strictly as a numerical perturbation experiment using the tabulated lower and upper scales, not as a reconstructed experiment-level covariance model. These are operational assumptions of the worked example, not evidence that all underlying systematic effects are independent.

\begin{table*}[t]
	\centering
	\caption{Neutron-lifetime data used for the controlled same-dataset comparison. Values and method labels follow Zhang et al. \cite{Zhang2022}; primary measurement papers are cited in the study column.}
	\label{tab:data}
	
	\footnotesize
	\renewcommand{\arraystretch}{1.08}
	\setlength{\tabcolsep}{4pt}
	
	\begin{tabularx}{\textwidth}{@{}Z{0.75cm} Y Z{1.7cm} Y Z{1.5cm} Z{2.3cm}@{}}
		\toprule
		Index & Study & Lifetime (s) & Quoted uncertainty & Type & PDG note \\
		\midrule
		
		1 &
		Ezhov 2018 \cite{Ezhov2018} &
		878.30 &
		$\pm1.6\ \pm1.0$ &
		Bottle &
		PDG \\
		
		2 &
		Serebrov 2018 \cite{Serebrov2018} &
		881.50 &
		$\pm0.7\ \pm0.6$ &
		Bottle &
		PDG \\
		
		3 &
		Pattie 2018 \cite{Pattie2018} &
		877.70 &
		$\pm0.7\,{}^{+0.4}_{-0.2}$ &
		Bottle &
		PDG \\
		
		4 &
		Leung 2016 \cite{Leung2016} &
		887.00 &
		$\pm39$ &
		Bottle &
		not in PDG \\
		
		5 &
		Arzumanov 2015 \cite{Arzumanov2015} &
		880.20 &
		$\pm1.2$ &
		Bottle &
		PDG \\
		
		6 &
		Yue 2013 \cite{Yue2013} &
		887.70 &
		$\pm1.2\ \pm1.9$ &
		Beam &
		PDG \\
		
		7 &
		Steyerl 2012 \cite{Steyerl2012} &
		882.50 &
		$\pm1.4\ \pm1.5$ &
		Bottle &
		PDG \\
		
		8 &
		Pichlmaier 2010 \cite{Pichlmaier2010} &
		880.70 &
		$\pm1.3\ \pm1.2$ &
		Bottle &
		PDG \\
		
		9 &
		Serebrov 2005 \cite{Serebrov2005} &
		878.50 &
		$\pm0.7\ \pm0.3$ &
		Bottle &
		PDG \\
		
		10 &
		Byrne 1996 \cite{Byrne1996} &
		889.20 &
		$\pm3.0\ \pm3.8$ &
		Beam &
		PDG \\
		
		11 &
		Mampe 1993 \cite{Mampe1993} &
		882.60 &
		$\pm2.7$ &
		Bottle &
		PDG \\
		
		12 &
		Alfimenkov 1990 \cite{Alfimenkov1990} &
		888.40 &
		$\pm2.9$ &
		Bottle &
		PDG not used \\
		
		13 &
		Kossakowski 1989 \cite{Kossakowski1989} &
		878.00 &
		$\pm27\ \pm14$ &
		Beam &
		PDG not used \\
		
		14 &
		Paul 1989 \cite{Paul1989} &
		877.00 &
		$\pm10$ &
		Bottle &
		PDG not used \\
		
		15 &
		Last 1988 \cite{Last1988} &
		876.00 &
		$\pm10\ \pm19$ &
		Beam &
		PDG not used \\
		
		16 &
		Spivak 1988 \cite{Spivak1988} &
		891.00 &
		$\pm9$ &
		Beam &
		PDG not used \\
		
		17 &
		Kosvintsev 1986 \cite{Kosvintsev1986} &
		903.00 &
		$\pm13$ &
		Bottle &
		PDG not used \\
		
		18 &
		Kosvintsev 1980 \cite{Kosvintsev1980} &
		875.00 &
		$\pm95$ &
		Bottle &
		PDG not used \\
		
		19 &
		Christensen 1972 \cite{Christensen1972} &
		918.00 &
		$\pm14$ &
		Beam &
		PDG not used \\
		
		20 &
		Gonzalez 2021 \cite{Gonzalez2021} &
		877.75 &
		$\pm0.28\,{}^{+0.22}_{-0.16}$ &
		Bottle &
		PDG \\
		
		21 &
		Wilson 2021 \cite{Wilson2021} &
		887.00 &
		$\pm14\,{}^{+7}_{-3}$ &
		Space &
		PDG not used \\
		
		\bottomrule
	\end{tabularx}
\end{table*}

All lifetime and uncertainty entries in Table~\ref{tab:data} are in seconds. The table contains \BottleCount{} bottle measurements, \BeamCount{} beam measurements, and \SpaceCount{} space-based result. The PDG-status strings are retained as compilation metadata rather than reinterpreted as an analysis variable; they are not used as an inclusion filter, and all 21 rows enter the calculations. This method heterogeneity is one reason the dataset is used here as an implementation benchmark rather than as a basis for a recommended pooled neutron lifetime.

\begin{figure*}[t]
\centering
\begin{minipage}[t]{0.49\textwidth}
\centering
\includegraphics[width=\linewidth]{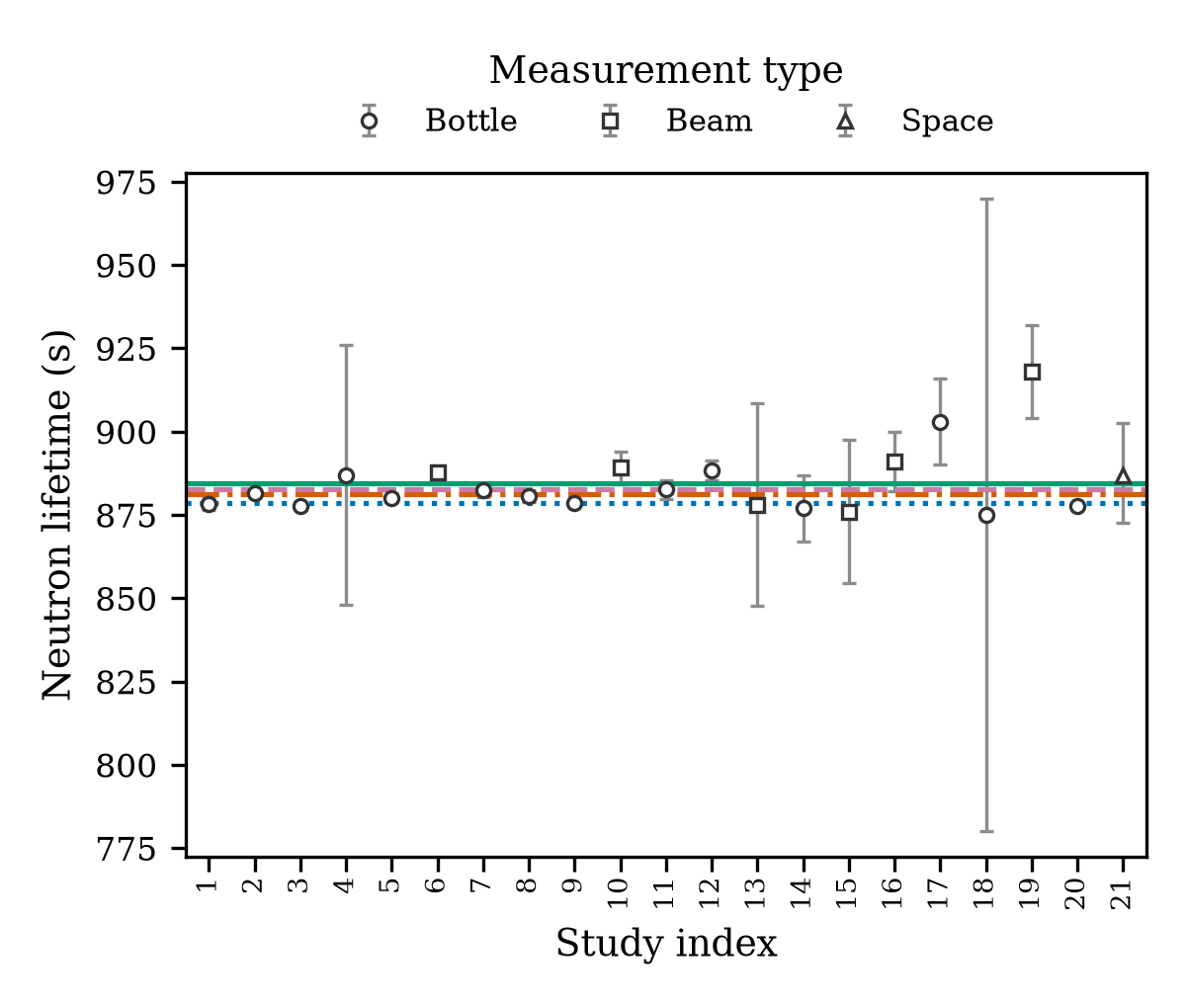}\\[-0.3em]
\small (a) Full reported range
\end{minipage}\hfill
\begin{minipage}[t]{0.49\textwidth}
\centering
\includegraphics[width=\linewidth]{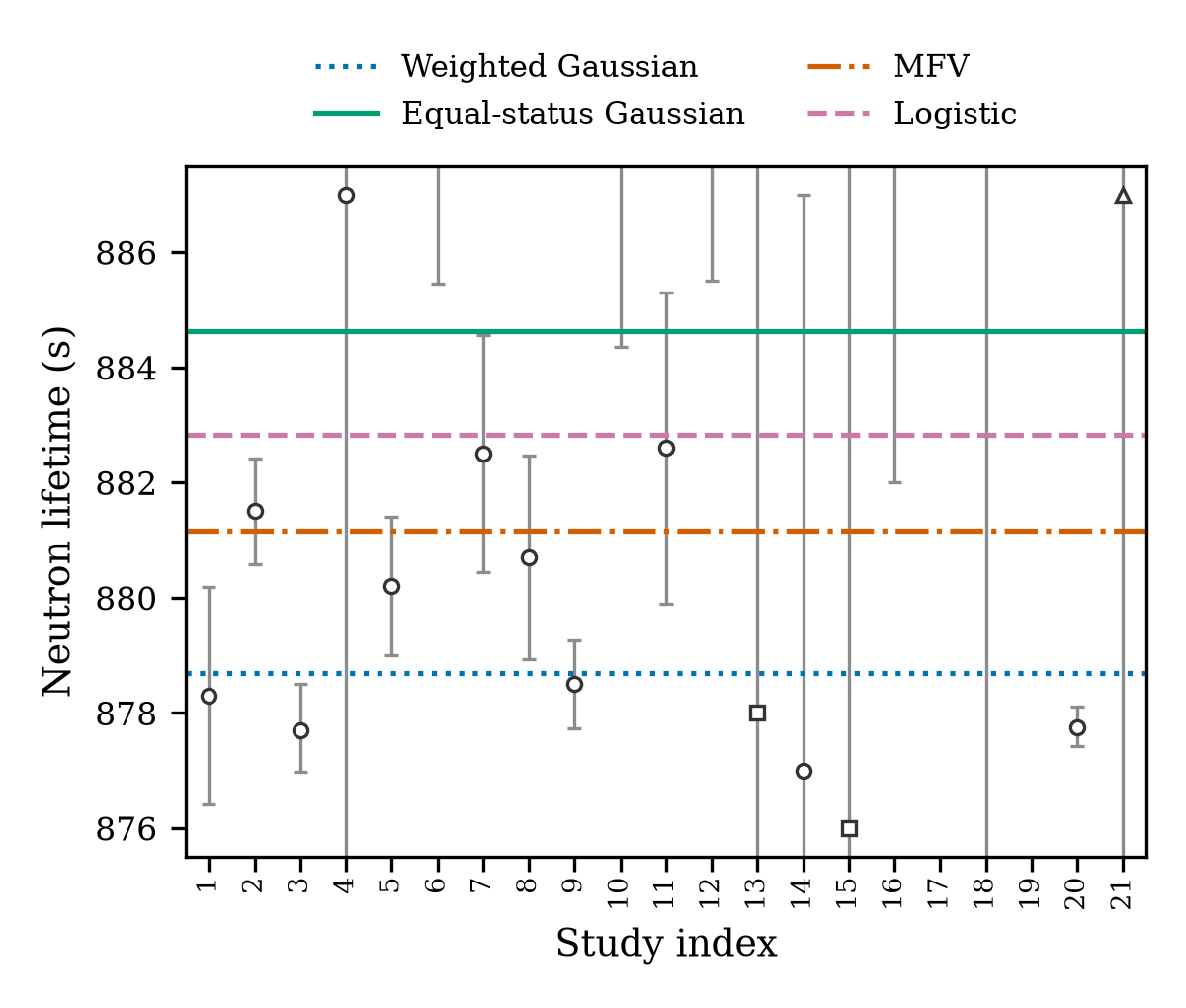}\\[-0.3em]
\small (b) Central zoom
\end{minipage}
\caption{The same 21 neutron-lifetime measurements with the inverse-variance weighted
	Gaussian, equal-status Gaussian, MFV, and logistic deterministic locations.
	Marker shapes distinguish bottle (circles), beam (squares), and space-based
	(triangles) measurements, while the location summaries use distinct line
	styles as well as colors. Error bars show the lower and upper uncertainty
	scales used in HPB, whereas the inverse-variance weighted calculation uses the symmetrized uncertainties defined in Section~\ref{sec:methods}. Study indices correspond to the first column of Table~\ref{tab:data}. Panel (a) retains the full
	reported uncertainty range, whereas panel (b) shows the central
	875.5--887.5~s region to make the separation among the four location
	estimates easier to see. Uncertainty bars extending outside this window are clipped by the zoom.}
\label{fig:dataest}
\end{figure*}

Figure~\ref{fig:dataest} places four deterministic location summaries on the same 21 measurements. The equal-status Gaussian mean provides a control that changes only the treatment of precision relative to the inverse-variance weighted Gaussian location. The MFV and logistic locations also give equal status to the central values but change the fixed-scale score shape: the MFV score redescends, whereas the logistic score remains bounded with a nonzero tail contribution.

\section{Results}
\label{sec:results}

This section reports the numerical results for the fixed 21-measurement
benchmark. We first compare the Gaussian controls, including the PDG/Birge
and residual-based (sandwich) uncertainty estimates. We then present the deterministic MFV
and logistic solutions together with their numerical diagnostics. Finally,
we compare the ordinary-bootstrap and HPB summaries and show how the
different uncertainty constructions change the reported ranges without
changing the underlying deterministic estimates.

We carried out the numerical calculations in Python using NumPy
\cite{Harris2020NumPy}, pandas \cite{McKinney2010Pandas}, and statsmodels
\cite{Seabold2010Statsmodels}. SciPy \cite{Virtanen2020SciPy} is retained in the pinned computational environment as a dependency. We used Matplotlib
\cite{Hunter2007Matplotlib} to create the figures.

\subsection{Gaussian controls, PDG/Birge, and residual-based uncertainty results}
The equal-status Gaussian control gives an arithmetic mean of \EqualGaussianMean~s and an empirical RMS scale of \EqualGaussianRMS~s.

Equation~\eqref{eq:weightedmean} gives an inverse-variance weighted mean of \WeightedMean~s with nominal internal standard uncertainty \InternalSE~s. For these 21 measurements, $\chi^2=\ChiSq$ and the reduced chi-square is \ReducedChiSq. Equation~\eqref{eq:pdgscale} gives $S_{\mathrm{PDG}}=\PDGScale$ and a PDG/Birge-style standard uncertainty of \PDGSE~s.

Equation~\eqref{eq:steiner_gauss_scale} gives the Gaussian dihesion factor $S_G=\SteinerGScale$. Under the mapping $V_G=S_G^2/W$, this quantity is a variance. Taking its square root gives the standard error \HCzeroSE~s, exactly equal to the HC0 residual-based standard error defined in Section~\ref{sec:methods}. A weighted least-squares software cross-check reproduces the direct HC0 standard-error calculation to machine precision and confirms the algebraic equivalence. The HC1--HC3 adjustments give \HConeSE~s for HC1, \HCtwoSE~s for HC2, and \HCthreeSE~s for HC3.

The largest leverage, meaning the largest fraction of the total inverse-variance weight carried by one measurement, is \MaxLeverage. Removing \MaxLOOStudy{} produces the largest weighted-mean leave-one-out change, \MaxLOOShift{}~s. These are related but distinct diagnostics because the deletion shift depends on both the measurement's weight fraction and how far its value lies from the fitted mean. The three largest inverse-variance weights account for \TopThreeWeightPct\% of the total weight. The weight-concentration effective count, $1/\sum_i h_i^2$, is \EffectiveWeightedN; it describes how concentrated the nominal weights are and should not be read as a count of independent physical constraints.

Table~\ref{tab:methoddiag} gives a compact comparison by measurement method. The within-method weighted means use the same fixed symmetrized uncertainties as the pooled weighted calculation; they are shown only to describe the benchmark's composition.

\begin{table}[t]
	\centering
	\caption{Comparison by measurement method for the neutron benchmark. The weight fraction gives each method's share of the total inverse-variance weight in the pooled Gaussian estimate.}
	\label{tab:methoddiag}
	\small
	\renewcommand{\arraystretch}{1.10}
	
	\begin{tabularx}{\columnwidth}{@{}l c C C C@{}}
		\toprule
		Method &
		$n$ &
		Equal-status mean (s) &
		Weighted mean (s) &
		Weight (\%) \\
		\midrule
		
		Bottle &
		\BottleCount &
		\BottleEqualMean &
		\BottleWeightedMean &
		\BottleWeightPct \\
		
		Beam &
		\BeamCount &
		\BeamEqualMean &
		\BeamWeightedMean &
		\BeamWeightPct \\
		
		Space &
		\SpaceCount &
		\SpaceEqualMean &
		\SpaceWeightedMean &
		\SpaceWeightPct \\
		
		\bottomrule
	\end{tabularx}
\end{table}

As an algebraic check, assigning the same uncertainty to every measurement
makes the HC1 and Birge/PDG uncertainties agree to machine precision,
confirming the limiting relation derived after
Equation~\eqref{eq:pdgscale}.

\subsection{MFV and logistic deterministic solutions}
With the initialization and stopping criterion specified in the Methods section, the deterministic MFV iteration gives $M_{\mathrm{MFV}}=\MFVvalue$~s with dihesion $\varepsilon=\MFVeps$~s. Zhang et al. reported 881.16~s, so the present single-path implementation reproduces their MFV benchmark \cite{Zhang2022}. The stated parameter-change criterion is first met at iteration 48; at that stopping point the maximum bounded equation residual is \MFVResidual. Continuing the same iterative procedure solely for the stricter numerical audit reaches a residual of \MFVAuditResidual{} at iteration 50. This diagnostic continuation does not replace the iteration-48 estimator.

The deterministic logistic iteration starts from the declared
$\mathrm{MAD}/\ln 3$ scale, $S^{(0)}=\LogisticInitScale$~s, and returns
$M_{\log}=\LogisticM$~s and $S_{\log}=\LogisticS$~s. At the stopping point,
the changes in $M$ and $S$ are \LogisticProdDeltaM~s and
\LogisticProdDeltaS~s, respectively, both below the $10^{-5}$~s stopping
threshold. The maximum equation residual is \LogisticResidual, indicating
numerical agreement to essentially machine precision. For the location
uncertainty, the ideal-logistic approximation gives
\LogisticIdealSE~s, compared with \LogisticSandSE~s from the residual-based
sandwich calculation.

\begin{figure*}[t]
\centering
\begin{minipage}[t]{0.48\textwidth}
\centering
\includegraphics[width=\linewidth]{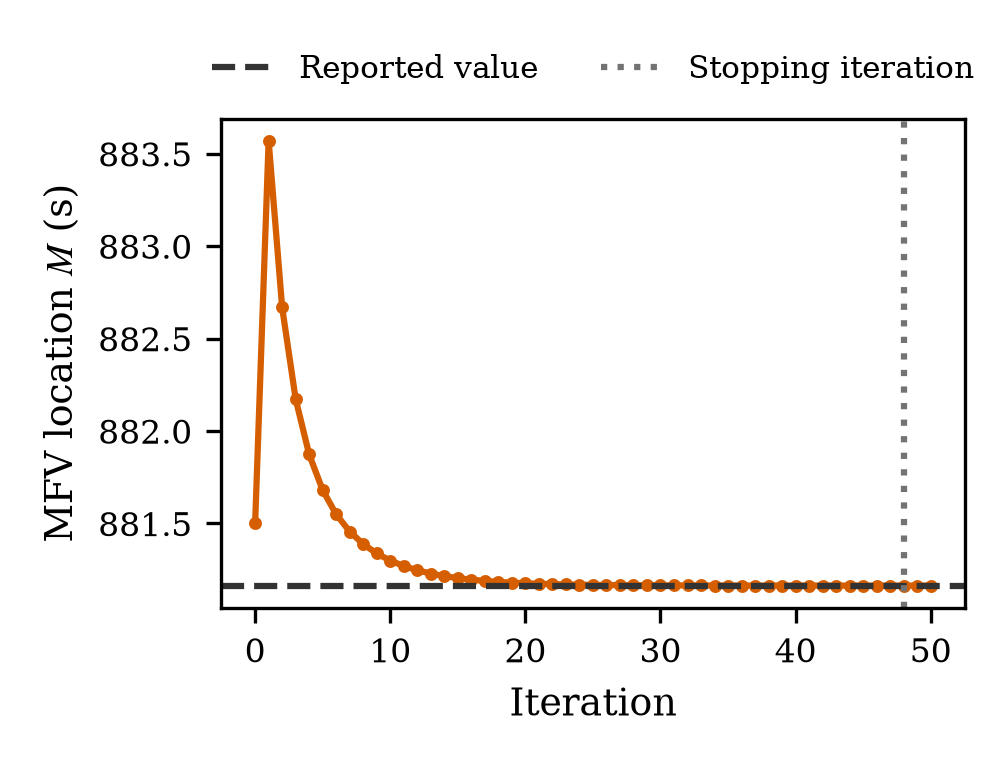}\\[-0.3em]
\small (a) MFV location
\end{minipage}\hfill
\begin{minipage}[t]{0.48\textwidth}
\centering
\includegraphics[width=\linewidth]{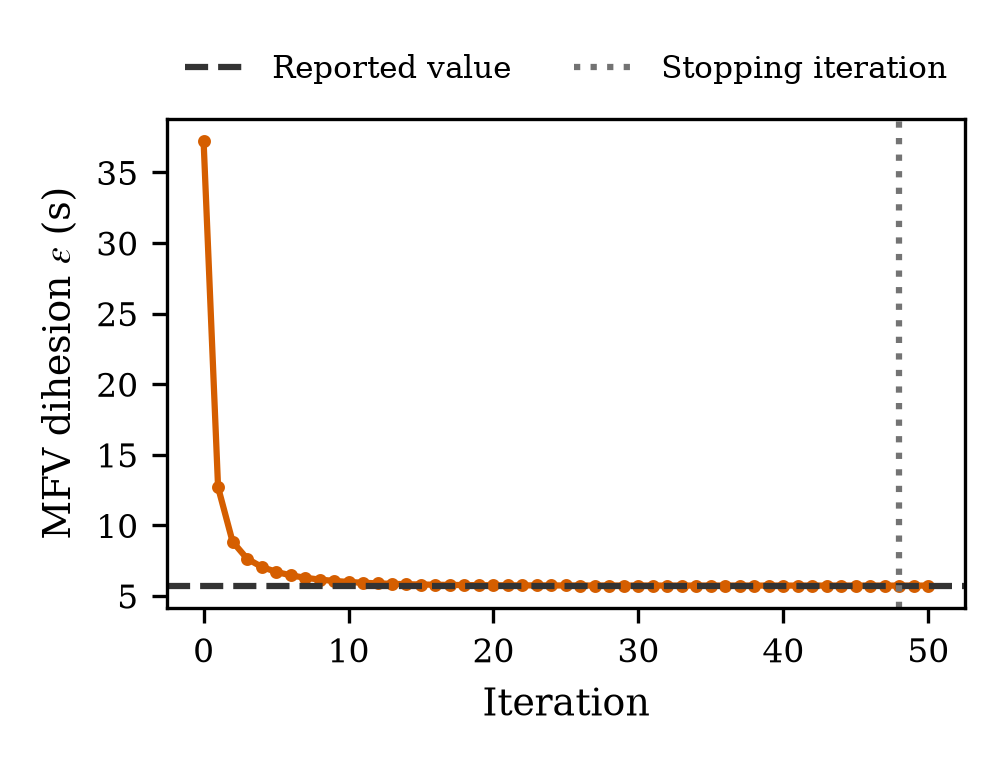}\\[-0.3em]
\small (b) MFV dihesion
\end{minipage}

\vspace{0.6em}
\begin{minipage}[t]{0.48\textwidth}
\centering
\includegraphics[width=\linewidth]{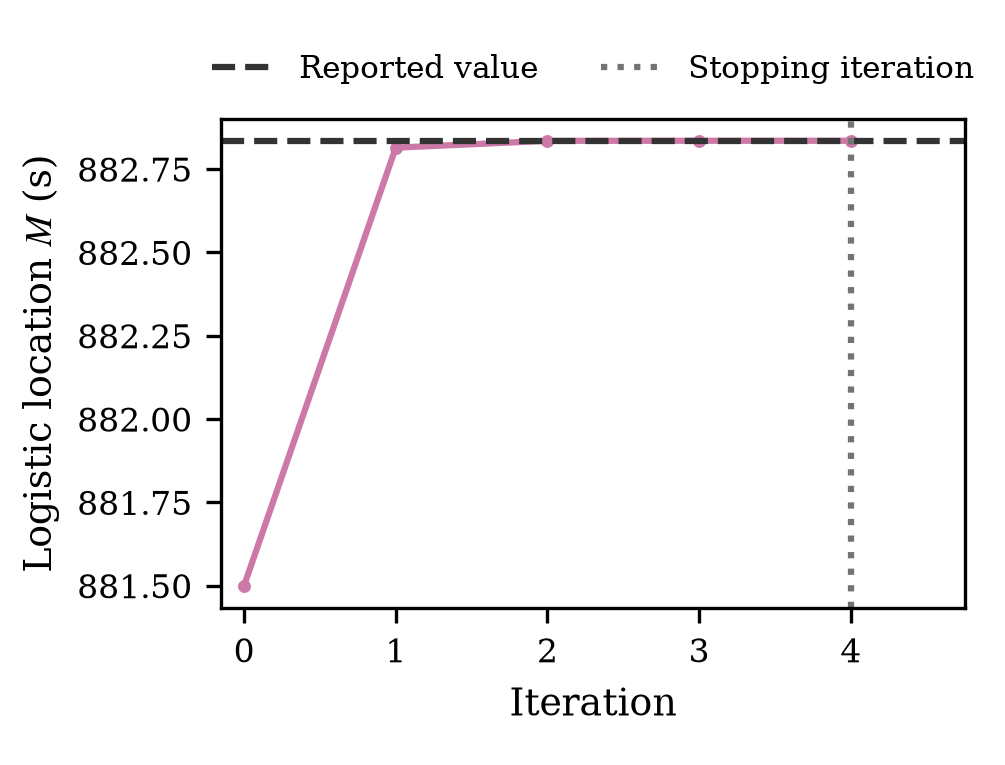}\\[-0.3em]
\small (c) Logistic location
\end{minipage}\hfill
\begin{minipage}[t]{0.48\textwidth}
\centering
\includegraphics[width=\linewidth]{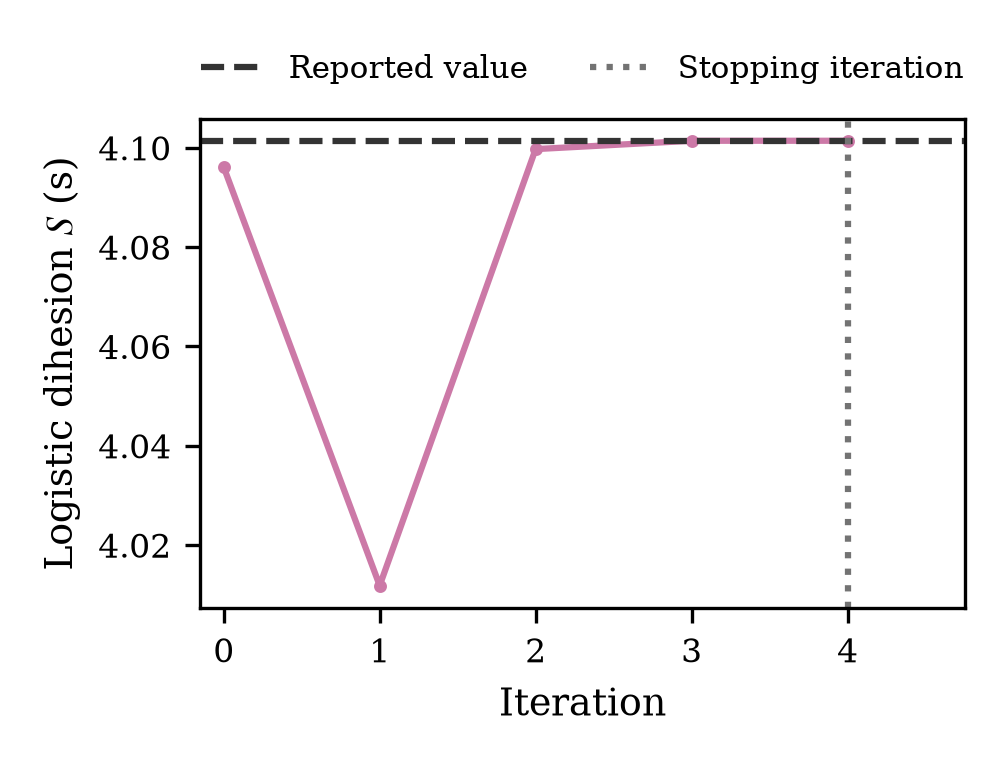}\\[-0.3em]
\small (d) Logistic dihesion
\end{minipage}
\caption{Iteration histories for the deterministic MFV and logistic calculations, including the initialized state. The vertical dotted line marks the stated stopping criterion. Points beyond that line, where present in the MFV panels, are optional numerical diagnostics and are not part of the estimator. Dashed horizontal lines mark the reported solutions.}
\label{fig:iterations}
\end{figure*}

Figure~\ref{fig:iterations} shows the deterministic iteration histories for the original 21-point dataset. The logistic estimates of $M$ and $S_{\log}$ reach the stated stopping criterion within only a few iterations, whereas the MFV estimates of $M$ and $\varepsilon$ require substantially more fixed-point updates.

\subsection{Bootstrap and HPB results}
The final resampling calculation uses \Nboot{} ordinary-bootstrap replicates and \Nboot{} HPB replicates. The MFV iteration returns a usable estimator value for every replicate in both calculations: \MFVBootValidN{} ordinary-bootstrap values and \MFVHPBValidN{} HPB values, with no unresolved MFV fits. Among these, \MFVBootZeroDihesionN{} ordinary-bootstrap replicates (\MFVBootZeroDihesionPct\%) and \MFVHPBZeroDihesionN{} HPB replicates (\MFVHPBZeroDihesionPct\%) reach the zero-scale limiting case defined in the Methods section. Of the 19,996 audit-applicable positive-scale MFV HPB outcomes, one is flagged by the stricter residual audit. That replicate remains valid and is included because the estimator is defined by the prescribed parameter-change stopping rule, while the residual continuation is diagnostic only. Logistic fits converge for all \Nboot{} ordinary-bootstrap and all \Nboot{} HPB replicates.

The central 68.27\% descriptive MFV ordinary-bootstrap percentile range is
\[
[\MFVBootLow,\MFVBootHigh]\ {\rm s},
\]
with estimated Monte Carlo standard errors of approximately \MFVBootMCSELow{}~s and \MFVBootMCSEHigh{}~s at the lower and upper endpoints. The central 68.27\% descriptive MFV HPB percentile range is
\[
[\MFVHPBLow,\MFVHPBHigh]\ {\rm s},
\]
with endpoint Monte Carlo standard errors of approximately \MFVHPBMCSELow{}~s and \MFVHPBMCSEHigh{}~s.

\begin{table*}[t]
	\centering
	\caption{Selected results for the controlled 21-point comparison. Analytical rows report one-standard-error ranges. Bootstrap and HPB rows report central 68.27\% descriptive percentile endpoints.}
	\label{tab:mainresults}
	\small
	\renewcommand{\arraystretch}{1.12}
	
	\begin{tabularx}{\textwidth}{@{}Y Z{2.7cm} C@{}}
		\toprule
		Estimator / summary
		& Point estimate (s)
		& Reported range or endpoints (s) \\
		\midrule
		\multicolumn{3}{@{}l}{\textit{Analytical standard-error summaries}} \\
		\addlinespace[0.2em]
		Weighted mean, internal SE
		& $\WeightedMean$
		& $[\InternalLow,\;\InternalHigh]$ \\
		
		Weighted mean, PDG/Birge SE
		& $\WeightedMean$
		& $[\PDGLow,\;\PDGHigh]$ \\
		
		Weighted mean, curvature/HC0 SE
		& $\WeightedMean$
		& $[\HCzeroLow,\;\HCzeroHigh]$ \\
		
		Weighted mean, HC3 SE
		& $\WeightedMean$
		& $[\HCthreeLow,\;\HCthreeHigh]$ \\
		
		Logistic, sandwich SE
		& $\LogisticM$
		& $[\LogisticSandLow,\;\LogisticSandHigh]$ \\
		
		\addlinespace
		\multicolumn{3}{@{}l}{\textit{Descriptive resampling summaries}} \\
		\addlinespace[0.2em]
		Weighted mean, ordinary bootstrap
		& $\WeightedMean$
		& $[\WBootLow,\;\WBootHigh]$ \\
		
		Weighted mean, HPB
		& $\WeightedMean$
		& $[\WHPBLow,\;\WHPBHigh]$ \\
		
		MFV, ordinary bootstrap
		& $\MFVvalue$
		& $[\MFVBootLow,\;\MFVBootHigh]$ \\
		
		MFV, HPB
		& $\MFVvalue$
		& $[\MFVHPBLow,\;\MFVHPBHigh]$ \\
		
		Logistic, ordinary bootstrap
		& $\LogisticM$
		& $[\LogBootLow,\;\LogBootHigh]$ \\
		
		Logistic, HPB
		& $\LogisticM$
		& $[\LogHPBLow,\;\LogHPBHigh]$ \\
		
		\bottomrule
	\end{tabularx}
\end{table*}

For the logistic location, the corresponding central 68.27\% descriptive ordinary-bootstrap percentile range is
\[
[\LogBootLow,\LogBootHigh]\ {\rm s},
\]
and the central 68.27\% descriptive HPB percentile range is
\[
[\LogHPBLow,\LogHPBHigh]\ {\rm s}.
\]
For reference, the weighted-mean central 68.27\% descriptive ordinary-bootstrap and HPB percentile ranges are
\[
[\WBootLow,\WBootHigh]\ {\rm s}
\quad\text{and}\quad
[\WHPBLow,\WHPBHigh]\ {\rm s},
\]
respectively. These are the central 68.27\% descriptive percentile ranges defined in the Methods section.

The deterministic MFV, $881.164$~s, rounds to the
$881.16$~s reported by Zhang et al.\ \cite{Zhang2022}.
The ordinary-bootstrap central 68.27\% percentile range obtained here,
$[878.911,\,883.428]$~s, also agrees closely with their reported
$[878.81,\,883.41]$~s range. The small differences in the bootstrap
endpoints are consistent with finite Monte Carlo resampling and do not
affect the agreement of the deterministic MFV values.

Table~\ref{tab:mainresults} summarizes the selected analytical
standard-error estimates and descriptive resampling ranges for the weighted,
MFV, and logistic results.

\begin{figure*}[t]
	\centering
\includegraphics[width=0.99\textwidth]{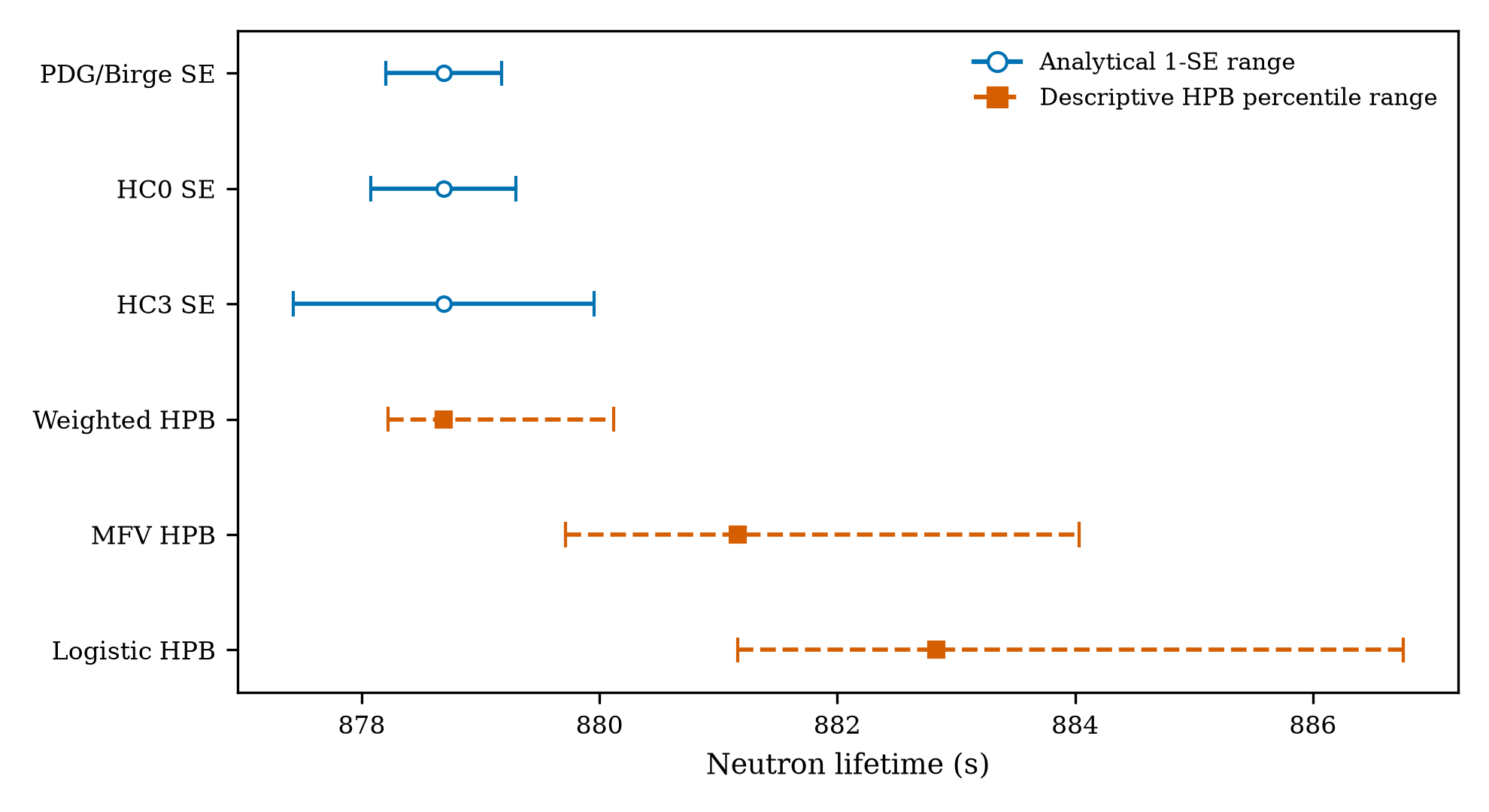}
	\caption{Selected summaries from the same 21-point dataset. Points mark the deterministic estimates. PDG/Birge, HC0, and HC3 are analytical standard-error constructions and are shown with open circles and solid ranges. Weighted, MFV, and logistic HPB entries are central 68.27\% descriptive percentile ranges and are shown with filled squares and dashed ranges; their repeated-sampling coverage is not calibrated.}
	\label{fig:selected_uncertainty}
\end{figure*}

Figure~\ref{fig:selected_uncertainty} provides a visual comparison of
selected analytical standard-error ranges and descriptive HPB percentile
ranges for the same dataset.

\section{Discussion}
\label{sec:discussion}

The Gaussian controls separate two effects that would otherwise be
confounded. Moving from the inverse-variance weighted Gaussian location to
the equal-status Gaussian mean removes the quoted precision weights while
retaining a linear score. Moving from the equal-status Gaussian mean to MFV
or logistic then changes the score shape while keeping equal status for the
reported central values. The resulting differences therefore should not be
described simply as a competition between ``nonrobust'' and ``robust''
estimators. The equal-status Gaussian RMS is a spread scale rather than a
standard error of the mean. For the weighted Gaussian result, the internal
standard error, PDG/Birge inflation, and HC0--HC3 covariance estimates answer
different uncertainty questions described in Section~\ref{sec:methods}; none
changes the weighted location itself.

When the underlying distribution is unknown, the choice of estimator should
follow the information available about the measurements rather than the
numerical result one prefers. If the quoted uncertainties provide meaningful
relative precision information, the Gaussian construction with unequal quoted
uncertainties uses that information directly through inverse-variance weighting. If the
reported central values should instead receive equal status, the equal-status
Gaussian mean provides a linear reference, while MFV and logistic reduce the
effect of large residuals in different ways. At fixed positive scale, the MFV
score redescends toward zero, whereas the logistic score remains bounded and
approaches a finite nonzero tail contribution. Neither tail response is
universally preferable. When the physical problem does not select one in
advance, comparing both as prespecified sensitivity summaries shows whether
the estimated location depends strongly on the assumed tail response and
avoids choosing a method after seeing the preferred numerical answer. These score properties describe the estimating equations; they do not by
themselves determine how the full joint location--scale procedures respond
to arbitrarily distant observations or how robust they remain when outliers
are present. If the measurements represent physically distinct populations, a single
pooled location estimate is not appropriate.

The Gaussian result for measurements with unequal quoted uncertainties has a narrower methodological meaning.
The experiment-level additive curvature balance is an adopted convention that
extends the single-density Steiner rule; neither the full joint product
density nor KL minimization uniquely implies it. Under this convention and
the additional mapping $V_G=S_G^2/W$, the resulting expression is
algebraically identical to the intercept-only HC0 variance. This identity
does not make HC0 a curvature-calibrated uncertainty. The fitted Gaussian
dihesion factor also should not be interpreted as a calibrated common
multiplier of the experimental uncertainties. In this dataset the maximum
leverage---the largest fraction of total inverse-variance weight carried by one
measurement---is \MaxLeverage, whereas the largest leave-one-out shift is
\MaxLOOShift{}~s. The weight fraction measures potential influence from quoted
precision, while the realized deletion shift also depends on the residual
through the identity given in Section~\ref{sec:methods}. This distinction
helps explain why HC2 and especially HC3, which give additional protection
against highly weighted measurements, exceed HC0.

The neutron compilation is intentionally treated as a benchmark rather than
as one homogeneous physical population. Bottle, beam, and space measurements
can have different systematic effects, and the available compilation does not
reconstruct all shared covariances or the meaning of every quoted uncertainty
component. The comparison by measurement method shows that bottle measurements carry most of the inverse-variance weight in the pooled Gaussian estimate. The weight-concentration effective count therefore describes how strongly the nominal precision weights are concentrated; it is not a count of independent physical constraints. A single pooled location cannot resolve differences among measurement methods. A new neutron-lifetime evaluation would require source-level
covariance information and a model that accounts for differences among
measurement methods.

The HPB and ordinary-bootstrap ranges have a deliberately descriptive
interpretation. Ordinary case resampling treats the study records as if they
were interchangeable draws from the empirical compilation, while HPB also
perturbs each selected study according to its quoted uncertainty. Because a
full model of between-study variation and shared correlations is unavailable,
these procedures produce method-dependent sensitivity distributions rather
than calibrated confidence procedures. We therefore do not interpret the central 68.27\% descriptive
percentile ranges as confidence intervals. The zero-dihesion component also places a discrete probability mass at zero rather than producing a completely smooth resampling distribution, which reinforces this distinction.

The logistic sandwich standard error provides a useful large-sample covariance
diagnostic because it combines the observed variability of the score
contributions with how strongly the coupled estimating equations respond to
changes in the fitted parameters. Its usual interpretation still requires the
fitted solution to lie away from parameter boundaries and the sample to be
large enough for the approximation to be reliable. With only 21 studies,
it therefore should not be interpreted as a small-sample coverage guarantee.

The numerical procedures provide operational definitions for this benchmark.
MFV and logistic use prescribed initializations, iterative schemes, and
application-scaled stopping criteria. Equation residuals evaluated at the
stopping point provide numerical diagnostics but do not define separate
solutions. Appendix~\ref{app:numerical} gives the implementation details and
special cases. Broader claims about bias, coverage, or solver reliability
would require dedicated simulations and comparisons beyond the scope of this
article.

\section{Conclusions}
\label{sec:conclusions}
The central organizing result is the separation of two operations: fixed-scale KL stationarity supplies the location equation, while a separate curvature rule defines its companion scale. MFV, logistic M-estimation, HPB, and sandwich covariance are established ingredients; the article-specific contribution is their placement, where relevant, inside one explicit Gaussian--Cauchy--logistic construction with a reproducible physics workflow.

The Gaussian cases provide conventional controls and, under the adopted experiment-level curvature convention plus the stated variance mapping, an exact algebraic connection to intercept-only HC0. The Cauchy case reproduces the established MFV reference. For the logistic substitute, the established bounded $\tanh$ location score is paired with the curvature equation $\langle\tanh^2[(x-M)/(2S)]\rangle=1/3$, which is distinct from scale-likelihood stationarity and from a separately tuned logistic M-scale.

On the fixed 21-measurement neutron-lifetime benchmark, the resulting deterministic locations and uncertainty diagnostics can be reproduced with one documented implementation. HPB central 68.27\% descriptive percentile ranges are reported only as sensitivity summaries, and no new recommended neutron lifetime is proposed. General properties such as bias and coverage, and the physical differences among experiment classes, remain separate questions for simulation and dedicated neutron-data evaluation.

\backmatter

\bmhead{Acknowledgements}

The author gratefully acknowledges Mariya Filimonova for her support,
which made it possible to devote the time needed to this work.

\bmhead{Funding}
The author declares that no funds, grants, or other support were received during the preparation of this manuscript.

\bmhead{Competing interests}
The author has no relevant financial or non-financial interests to disclose.

\bmhead{Author contributions}
The sole author was responsible for the conception and design of the study, methodology, analysis, software, interpretation of the results, and preparation and revision of the manuscript.

\bmhead{Data and code availability}
The data and code supporting the findings of this study will be made publicly available in an open repository upon acceptance of the manuscript for publication. A persistent link to the repository will be provided in the final published version. The complete analysis package is available to the editors and reviewers upon request during peer review.

\begin{appendices}
\renewcommand{\theequation}{\thesection.\arabic{equation}}
\renewcommand{\thetable}{\thesection.\arabic{table}}
\renewcommand{\thefigure}{\thesection.\arabic{figure}}
\renewcommand{\theHequation}{app.\thesection.\arabic{equation}}
\renewcommand{\theHtable}{app.\thesection.\arabic{table}}
\renewcommand{\theHfigure}{app.\thesection.\arabic{figure}}

\section{Scale and uncertainty details}
\label{app:uncertainty}

This appendix collects formulas that support the main-text interpretation but are not needed to follow the derivation of the three location--scale constructions.

\subsection{Scale and uncertainty notation}

The scale symbol $S$ is used locally when a substitute distribution is introduced, but the method-specific quantities have different meanings. The word \emph{dihesion} is historical MFV terminology for the Cauchy scale; its Gaussian and logistic uses in this manuscript are explicit extensions to the analogous curvature-defined scales. Table~\ref{tab:scale_notation} summarizes the notation used when the methods are compared.
\begin{table*}[htbp]
	\centering
	\caption{Interpretation of the principal scale and uncertainty quantities used in the analysis. A curvature-defined dihesion describes the scale of the corresponding location--scale construction; it is not automatically a standard error of the fitted location.}
	\label{tab:scale_notation}
	
	\small
	\renewcommand{\arraystretch}{1.15}
	
	\begin{tabularx}{\textwidth}{@{}L{3.6cm}Y L{2.5cm}@{}}
		\toprule
		Symbol & Meaning & Units \\
		\midrule
		
		$\sigma_i$ &
		quoted standard uncertainty of measurement $i$ &
		units of $x$ \\
		
		$S$ (equal-status Gaussian) &
		empirical RMS residual scale &
		units of $x$ \\
		
		$S_G$ &
		Gaussian curvature/dihesion factor &
		dimensionless \\
		
		$V_G(M_w)=S_G^2/W$ &
		variance associated with the Gaussian curvature construction; algebraically equal to HC0 for the intercept-only weighted fit &
		units of $x^2$ \\
		
		$S_{\mathrm{PDG}}$ &
		PDG/Birge scale factor applied to the uncertainty of the weighted mean &
		dimensionless \\
		
		$\varepsilon$ &
		Cauchy/MFV dihesion &
		units of $x$ \\
		
		$S_{\log}$ &
		logistic dihesion &
		units of $x$ \\
		
		$1/\sqrt{W}$ &
		nominal internal standard error of the inverse-variance weighted mean &
		units of $x$ \\
		
		$\operatorname{SE}(M)$ &
		standard error of an estimated location &
		units of $x$ \\
		
		\bottomrule
	\end{tabularx}
\end{table*}
The repeated use of $S$ inside the Gaussian and logistic subsections is local notation for the scale parameter of the substitute currently under discussion. Cross-method comparisons use the qualified symbols in the table.

\subsection{HC0--HC3 for the weighted location}

For the weighted fit of one common constant, define $W=\sum_iw_i$, residuals $e_i=x_i-M_w$, and the weight fractions $h_i$. In regression terminology these fractions are called leverage:
\[
h_i=\frac{w_i}{W}.
\]
The HC0--HC3 variances are
\[
\begin{aligned}
\widehat V_{\mathrm{HC0}}(M_w)
&=\frac{1}{W^2}\sum_i w_i^2e_i^2,\\
\widehat V_{\mathrm{HC1}}(M_w)
&=\frac{n}{n-1}\widehat V_{\mathrm{HC0}}(M_w),\\
\widehat V_{\mathrm{HC2}}(M_w)
&=\frac{1}{W^2}\sum_i\frac{w_i^2e_i^2}{1-h_i},\\
\widehat V_{\mathrm{HC3}}(M_w)
&=\frac{1}{W^2}\sum_i\frac{w_i^2e_i^2}{(1-h_i)^2}.
\end{aligned}
\]
HC0 uses the raw squared residual contributions. HC1 multiplies HC0 by $n/(n-1)$, a finite-sample correction for estimating one common location from the same $n$ measurements. HC2 and HC3 increasingly enlarge the contribution from measurements that carry a large fraction of the total weight \cite{White1980,MacKinnonWhite1985}. These are variance estimators for the fitted weighted location. They do not convert the reported $\sigma_i$ into a calibrated common experimental-width model.

\subsection{Logistic sandwich covariance}

At the fitted logistic solution let $\widehat{\bm{\theta}}=(\widehat M,\widehat S)^T$ and
\[
\bm{\psi}_i(\bm{\theta})
=
\begin{pmatrix}
t_i\\
t_i^2-1/3
\end{pmatrix},
\qquad
t_i=\tanh\!\left(\frac{x_i-M}{2S}\right).
\]
Define
\[
A
=
\frac{1}{n}\sum_i
\left.
\frac{\partial\bm{\psi}_i}{\partial\bm{\theta}^T}
\right|_{\widehat{\bm{\theta}}},
\qquad
B
=
\frac{1}{n}\sum_i
\bm{\psi}_i(\widehat{\bm{\theta}})
\bm{\psi}_i(\widehat{\bm{\theta}})^T.
\]
The matrix $A$ measures how strongly the estimating equations change when $M$ or $S$ changes, while $B$ measures how much the individual score contributions vary across the observed measurements. The name \emph{sandwich covariance} refers to the matrix $B$ being placed between inverse sensitivity matrices in the expression below. In a one-parameter analogy, the same idea is score variability divided by squared score sensitivity; the matrix formula is the corresponding two-parameter calculation. The covariance estimate is
\[
\begin{aligned}
\widehat{\mathrm{Cov}}(\widehat{\bm{\theta}})
&=\frac{1}{n}A^{-1}BA^{-T},\\
\operatorname{SE}_{\mathrm{sand}}(M)
&=
\sqrt{
\left[\widehat{\mathrm{Cov}}(\widehat{\bm{\theta}})\right]_{11}
}.
\end{aligned}
\]
This construction does not require an exactly logistic parent distribution, but its usual interpretation still requires the fitted solution to lie away from a parameter boundary and the sensitivity matrix $A$ to be invertible (nonsingular). It also relies on a large-sample setting in which the study-level contributions satisfy the independence assumptions of the calculation. With only 21 studies, it should be treated as a large-sample covariance diagnostic rather than as evidence of calibrated small-sample coverage.

\section{Resampling workflow and reporting conventions}
\label{app:resampling}

The ordinary-bootstrap and HPB calculations use the same end-to-end workflow.
\begin{enumerate}
\item Draw $n$ row indices with replacement from the 21-study table. The selected central value, quoted lower and upper uncertainty scales, and study metadata move together as one study record.
\item For the ordinary bootstrap, leave each selected central value unchanged. For HPB, perturb it according to Equation~\eqref{eq:hpb}, while retaining the uncertainty fields associated with the selected study.
\item Recalculate the estimator with exactly the same deterministic definition used for the original dataset: inverse-variance weighting for the Gaussian location, the prescribed single-path MFV iteration for MFV, and the same prescribed damped-Newton iteration for the logistic estimator.
\item Record the estimator value and terminal status. If an MFV outcome satisfies the stopping criterion with raw $\varepsilon\le10^{-5}$~s, report $\varepsilon=0$ and retain it as a valid zero-scale limiting case. Equation residuals at the stopping point are recorded as numerical diagnostics. A genuinely unresolved fit is never silently discarded.
\item If every replicate is usable, report the 0.15865 and 0.84135 empirical quantiles as the central 68.27\% descriptive percentile range. If any replicate is unresolved, withhold the corresponding published range and retain conditional summaries only as diagnostics.
\end{enumerate}

The quantiles use NumPy's linear convention. Endpoint Monte Carlo error is approximated by perturbing the order-statistic rank by one binomial standard deviation. The calculation uses 20,000 replicates, bootstrap seed 20260825, HPB seed 20260826, and a fixed random-design block, so changing the numerical processing batch size does not change the generated resamples. The resampling outputs are sensitivity distributions under these rules.

\section{Numerical implementation and special cases}
\label{app:numerical}

This appendix gives the numerical details needed to reproduce the MFV and
logistic calculations. We first state the MFV stopping rule and treatment of
the zero-scale limit. We then describe the exact two-support logistic case
and the damped-Newton implementation used for the general logistic fit. These
details define the numerical conventions used throughout the analysis but do
not introduce separate estimators from those reported in the main text.

\subsection{MFV stopping criterion and zero-scale limit}
The MFV estimator is defined by the initialization and single iterative scheme given in the main text. The iteration stops when the changes in both $M$ and $\varepsilon$ are below $10^{-5}$~s. This threshold is far below the precision relevant to the reported result and the uncertainty scale of the input measurements. If the stopping criterion is satisfied with raw $\varepsilon\le10^{-5}$~s, the fitted scale is numerically indistinguishable from zero at the resolution adopted here; we therefore report $\varepsilon=0$. We refer to this only as the zero-scale limiting case. The raw numerical scale is retained in the diagnostics.

For a positive-scale reported solution, define
\[
\zeta_i=\frac{x_i-M}{\varepsilon},
\qquad
\omega_i=\frac{1}{1+\zeta_i^2},
\]
and
\[
R_M
=
\frac{\displaystyle\sum_i \omega_i(x_i-M)}
{s_{\mathrm{ref}}\displaystyle\sum_i \omega_i},
\]
where
\[
s_{\mathrm{ref}}
=
\max\!\left\{
x_{\max}-x_{\min},\,
\operatorname{sd}(x),\,
1~{\rm s}
\right\},
\]
together with
\[
r_\varepsilon
=
3\,
\frac{\displaystyle\sum_i(\zeta_i\omega_i)^2}
{\displaystyle\sum_i\omega_i^2},
\qquad
R_\varepsilon
=
\frac{r_\varepsilon-1}{r_\varepsilon+1}.
\]
For finite $\varepsilon>0$, these residuals have the same zeros as the MFV location and curvature equations while remaining numerically well scaled. We evaluate $\max(|R_M|,|R_\varepsilon|)$ at the stopping point as a numerical diagnostic. The residual diagnostic does not define the estimator, does not alter the reported $M$ or $\varepsilon$, and is not applied when $\varepsilon$ is reported as zero under this limiting rule.

The iteration allows at most $10^5$ steps. An exactly constant sample is returned directly at its common value with zero dihesion. A non-finite update, a numerical zero reached before the parameter-change stopping criterion, or exhaustion of the iteration cap before that criterion is met is classified as unresolved. The initialization and iteration together define the numerical estimator used here; the calculation reports the pair reached by this prescribed scheme and does not attempt to enumerate alternative stationary roots.

\subsection{Exact two-support logistic case}

Consider an exact two-value sample supported on $a$ and $b$ with multiplicities $n_a$ and $n_b$. Let
\[
t_a=\tanh\!\left(\frac{a-M}{2S}\right),
\qquad
t_b=\tanh\!\left(\frac{b-M}{2S}\right).
\]
Equations~\eqref{eq:loglocation} and \eqref{eq:logscale} imply
\[
n_a t_a+n_b t_b=0,
\qquad
\frac{n_a t_a^2+n_b t_b^2}{n_a+n_b}=\frac13.
\]
Eliminating one score gives
\[
t_a^2=\frac{n_b}{3n_a},
\qquad
t_b^2=\frac{n_a}{3n_b}.
\]
A finite positive scale requires $|t_a|<1$ and $|t_b|<1$, which yields Equation~\eqref{eq:logistic_two_cluster_domain}. At the limiting ratios $3{:}1$ and $1{:}3$, one required score reaches unit magnitude only as $S\to0$; a more imbalanced exact two-support sample therefore has no finite positive-scale solution.

\subsection{Logistic damped-Newton implementation}

For $z_i=(x_i-M)/(2S)$ define $t_i=\tanh z_i$ and $c_i=1-t_i^2=\sech^2z_i$. The Jacobian of Equations~\eqref{eq:loglocation} and \eqref{eq:logscale} is built from
\[
\frac{\partial \Psi_1}{\partial M}=-\frac{\langle c\rangle}{2S},
\qquad
\frac{\partial \Psi_1}{\partial S}=-\frac{\langle zc\rangle}{S},
\]
\[
\frac{\partial \Psi_2}{\partial M}=-\frac{\langle tc\rangle}{S},
\qquad
\frac{\partial \Psi_2}{\partial S}=-\frac{2\langle ztc\rangle}{S}.
\]
The damped Newton update is
\begin{equation}
\begin{aligned}
\bm{\theta}^{(k+1)}
&=
\bm{\theta}^{(k)}
-\lambda_kJ^{-1}\bm{\Psi}^{(k)},\\
\bm{\theta}
&=(M,S)^T,\\
\bm{\Psi}
&=(\Psi_1,\Psi_2)^T,
\qquad
0<\lambda_k\le1.
\end{aligned}
\label{eq:lognewton}
\end{equation}

The location initializer is the sample median. If
\[
\mathrm{MAD}
=
\operatorname{median}_i\left|x_i-\operatorname{median}(x)\right|
\]
is positive, the scale initializer is
\[
S^{(0)}=\frac{\mathrm{MAD}}{\ln3}.
\]
For a tied sample with $\mathrm{MAD}=0$ but nonzero spread, the usual
$\mathrm{MAD}/\ln 3$ initialization cannot provide a positive starting
scale. We therefore use
\[
S^{(0)}
=
\max\!\left[
\frac{\sqrt3}{\pi}\operatorname{sd}(x),
10S_{\mathrm{floor}}
\right].
\]
Here $\operatorname{sd}(x)$ is the population-form standard deviation of the observed sample, so it uses denominator $n$ rather than $n-1$. The factor $\sqrt3/\pi$ converts this standard deviation to the
scale parameter of a logistic distribution, for which
$\operatorname{sd}(X)=\pi S/\sqrt3$. The second term keeps the starting
value safely above the numerical scale floor.

We define a reference scale from the spread of the data as
\[
s_{\mathrm{ref}}
=
\max\!\left\{
x_{\max}-x_{\min},\,
\operatorname{sd}(x),\,
1~{\rm s}
\right\},
\]
and set
\[
S_{\mathrm{floor}}
=
\max\!\left(
10^{-10}s_{\mathrm{ref}},
10^{-12}~{\rm s}
\right).
\]
Here $S_{\mathrm{floor}}$ is a numerical safeguard, not a fitted scale
or a physical uncertainty. The relative term makes the safeguard follow
the scale of the data, while the absolute term prevents it from becoming
arbitrarily small. The $1~{\rm s}$ lower bound in $s_{\mathrm{ref}}$
provides a stable reference for data whose spread is already very small.

If the sample range does not exceed $S_{\mathrm{floor}}$, we treat the
sample as numerically indistinguishable from a constant sample at the
adopted resolution and do not attempt a positive-scale logistic fit.

An accepted residual-decreasing damped-Newton step satisfies the stopping criterion when
\[
|\Delta M|<10^{-5}~{\rm s},
\qquad
|\Delta S|<10^{-5}~{\rm s}.
\]
The application-specific unit is explicit: these thresholds define the neutron-lifetime implementation and should be transformed consistently if the data are expressed in another unit. At the stopping point we record
\[
R_{\log}=\max(|\Psi_1|,|\Psi_2|)
\]
as a numerical diagnostic; it does not define a second solution or change resampling inclusion. If the current state is already a floating-point root at machine-zero residual, it is recorded directly as a zero-change solution rather than being sent to a strict-decrease line search. Otherwise, if a full Newton step fails to reduce $R_{\log}$ or gives $S\le S_{\mathrm{floor}}$, the step is halved, with at most 32 halvings. The Jacobian condition number must not exceed $10^{14}$ and the hard iteration cap is 500. A singular or non-finite Jacobian, line-search failure, or exhaustion of the iteration cap before the parameter-change stopping criterion is met is a numerically unresolved outcome. Such a failure is not labeled ``no root.'' The stated initialization and damped-Newton iteration define the numerical solution reported here; no enumeration of alternative roots is required for this application.

\end{appendices}

\bibliography{references}%
\end{document}